\documentclass[aip,amsmath,amssymb,reprint]{revtex4-1}
\draft
\usepackage{graphicx}  
\usepackage{dcolumn}
\usepackage{amsmath}
\usepackage{makecell}
\usepackage{bm}        
\usepackage{amssymb}   

\usepackage[utf8]{inputenc}
\usepackage[T1]{fontenc}
\usepackage{mathptmx}

\usepackage{float}
\usepackage{color}

\begin{document}

\title{Machine learning short-ranged many-body interactions in colloidal systems using descriptors based on Voronoi cells}

\author{Rinske M. Alkemade}
\affiliation{Soft Condensed Matter and Biophysics, Debye Institute for Nanomaterials Science, Utrecht University, Utrecht, Netherlands }
\author{Rastko Sknepnek}
\affiliation{School of Life Sciences, University of Dundee, Dundee, DD1 5EH, United Kingdom}
\affiliation{School of Science and Engineering, University of Dundee, Dundee, DD1 4HN, United Kingdom}
\author{Frank Smallenburg}
\affiliation{
Universit\'e Paris-Saclay, CNRS, Laboratoire de Physique des Solides, 91405 Orsay, France
}
\author{Laura Filion}
\affiliation{Soft Condensed Matter and Biophysics, Debye Institute for Nanomaterials Science, Utrecht University, Utrecht, Netherlands }
\begin{abstract}

Machine learning (ML) strategies are opening the door to faster computer simulations, allowing us to simulate more realistic colloidal systems. Since the interactions in colloidal systems are often highly many-body, stemming from e.g. depletion and steric interactions, one of the challenges for these algorithms is capturing the many-body nature of these interactions. In this paper, we introduce a new ML-based strategy for fitting many-body interactions in colloidal systems where the many-body interaction is highly local. To this end, we develop Voronoi-based descriptors for capturing the local environment and fit the effective potential using a simple neural network. To test this algorithm, we consider a simple two-dimensional model for a colloid-polymer mixture, where the colloid-colloid interactions and colloid-polymer interactions are hard-disk like, while the polymers themselves interact as ideal gas particles. We find that a Voronoi-based description is sufficient to accurately capture the many-body nature of this system. Moreover, we find that the Pearson correlation function alone is insufficient to determine the predictive power of the network emphasizing the importance of additional metrics when assessing the quality of ML-based potentials.

\end{abstract}

\maketitle

\section{Introduction}
\begin{figure}[t]
    \centering
    \includegraphics[width=0.5\textwidth]{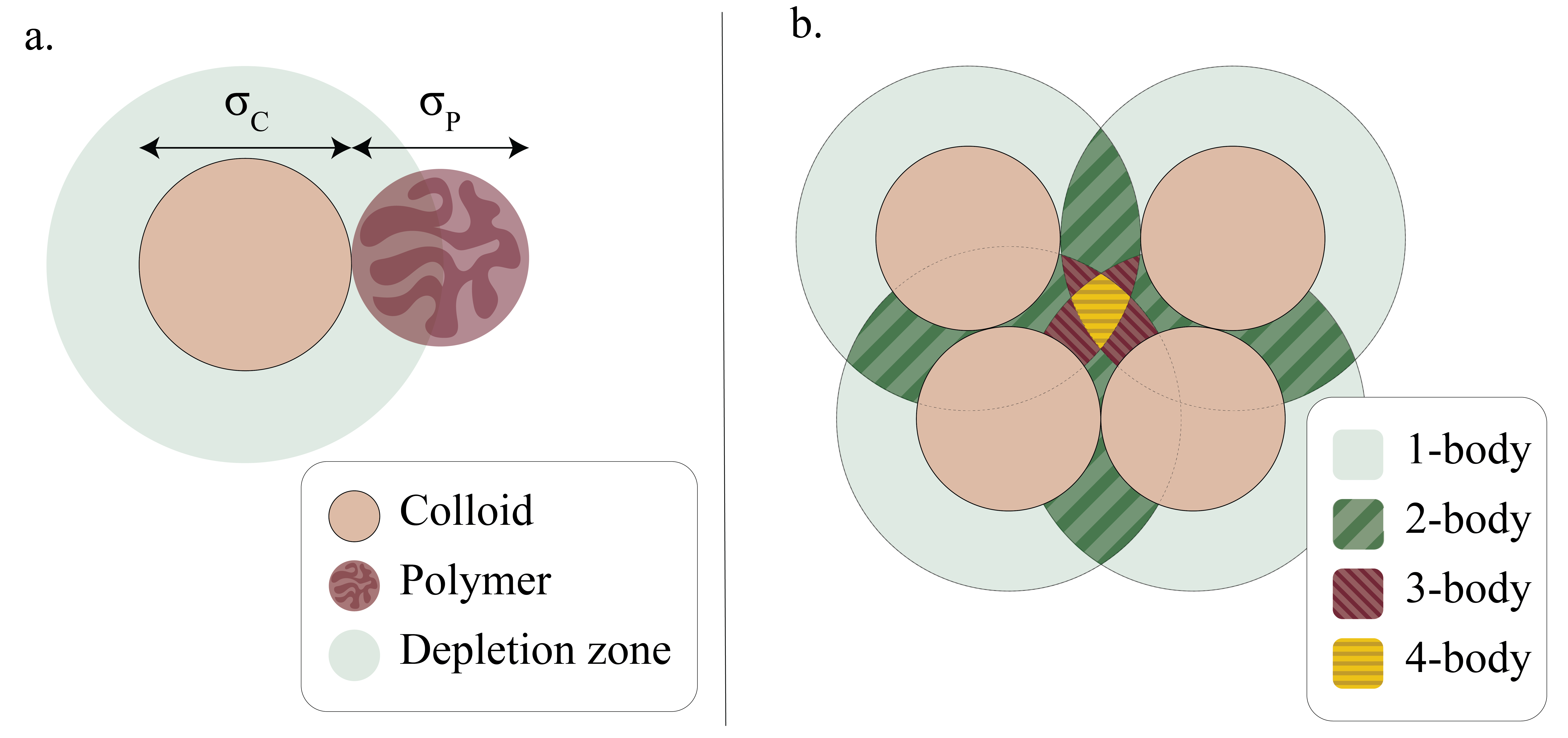}
    \caption{a. Cartoon of the 2D model system consisting of colloids and polymers. The light blue area indicates the depletion zone. b. Cartoon of how overlapping depletion zones can lead to multi-body interactions (here shown up to a 4-body interaction).}
    \label{fig:Depletion}
\end{figure}

Over the last few decades, progress in colloidal particle synthesis has led to an incredible collection of colloidal building blocks ranging from a few nanometers to microns in size, and with an ever-increasing variety of shapes and complex interactions. Together with these impressive strides in colloidal synthesis, there has been an increasing desire to be able to simulate such systems accurately enough to predict their phase behavior, as well as to target their design. However, accurately modeling colloidal systems can be very computationally costly as it requires modeling the colloid core, ligands on their surface, the solvent molecules in which the colloids are dispersed, polymers in the solvent, etc.  As such, even accurately simulating a small number of such particles can be computationally expensive. One approach to circumvent these expensive simulations is to integrate out some degrees of freedom associated with the system. This leads to an effective Hamiltonian that is exact, but often includes many-body terms. While such a Hamiltonian can in principle be used to capture the system's behaviour, the practical evaluation of these higher-order interactions can remain a significant computational bottleneck\cite{dijkstra2006effect, kobayashi2019correction}. 

One of the possible solutions to speed up the evaluation of these many-body terms -- and simulations in general -- lies in the realm of machine learning (ML). Over the past decade, ML has emerged as an important tool in speeding up computer simulations of atomic and molecular systems \cite{behler2016perspective, noe2020machine} by replacing expensive density functional theory calculations for electronic structure on the atomic scale, with machine learned forces and/or potentials. After training, these ML-based interaction potentials are exploited in either molecular dynamics or Monte Carlo (MC) simulations, allowing researchers to access larger system sizes and longer time scales than would be accessible otherwise. For such systems, strategies are rapidly evolving and improving -- from the early ``standard'' neural network (NN) approaches, to kernel fitting, on the fly training, and message passing neural networks 
\cite{behler2007generalized, artrith2016implementation,li2015molecular, jacobsen2018fly, jinnouchi2019phase, kocer2022neural, batatia2022mace}.  

Interestingly, however, on the colloidal scale, only a few studies have thus far explored the application of ML-based algorithms for speeding up simulations of systems with computationally expensive interactions. These include, e.g. the interaction between elastic spheres \cite{boattini2020modeling}, the effective interaction between nanoparticles \cite{chintha2021modeling, giunta2023coarse}, the effective interaction between colloids in the presence of depletants \cite{campos2021machine}, the interaction between anisotropic colloids \cite{campos2022machine, campos2024machine,  argun2024molecular}, and the interaction between nanoparticles with a single polymer chain grafted onto the surface \cite{gautham2022deep}. In most of these cases (with the exception of Ref. \onlinecite{gautham2022deep}), the choice of descriptors and machine learning techniques only allowed the ML-strategy to capture 2- and possibly 3-body contributions to the effective interaction. As a result, these methods are not well suited to deal with interactions that strongly depend on higher-order many-body terms.

A possible route forward in addressing the many-body nature in colloidal systems is to follow the advances made on the atomic scale and e.g. incorporate message passing neural networks such as MACE \cite{batatia2022mace}. Another option, that we explore here, is to develop alternative ways of capturing the many-body contributions from the structure. 

Two of the most important causes of many-body interactions in colloidal systems arise due to steric interactions and depletion interactions. These two types of interactions have one important feature in common -- they are highly local. Here, we want to explore whether we can exploit this short-ranged nature of colloidal depletion and steric many-body interactions to develop physically inspired structural descriptors that inherently capture all relevant many-body effects. To do this, we can take inspiration from phenomenological models, which incorporate many-body interactions, such as the models commonly used in modeling cell tissues and soft colloids. In particular, a commonly used approach here is a Voronoi-like description for the system  (e.g.\ Refs.\ \onlinecite{bi2016motility, ziherl2000soap}). Such a Voronoi description is naturally many-body and short-ranged, making it an ideal starting point for building new descriptors for e.g. depletion interactions.

Here, we examine a simple model system for colloidal particles with depletants and fit the effective interactions between the colloids using a Voronoi-based descriptions combined with a small neural network.  Note that this system is ideal as a test case as i) training data can be rapidly generated, and ii) the full model can be relatively easily simulated in order to confirm that the resulting ML-based simulations reproduce the correct behaviour.

We find that for a single size ratio (colloid-to-polymer diameter), a single Voronoi-inspired ML model can reproduce the structure of all relevant phases (gas, liquid, crystal) over all densities. Moreover, we find that the virial pressures in the ML system match the associated brute-force pressures. Additionally, we find that it is difficult to predict the performance of the ML models based on only the correlation between predicted and true effective interactions. In particular, high Pearson correlation coefficients ($>0.99999$) are insufficient to guarantee that the ML-based simulation will accurately reproduce the correct structure of the system. Hence, in this paper, we compare the radial distribution function of the fitted model to results from simulations where the polymers are treated explicitly to assess the quality of the potential.

\begin{figure}
    \centering
    \includegraphics[width=0.99\columnwidth]{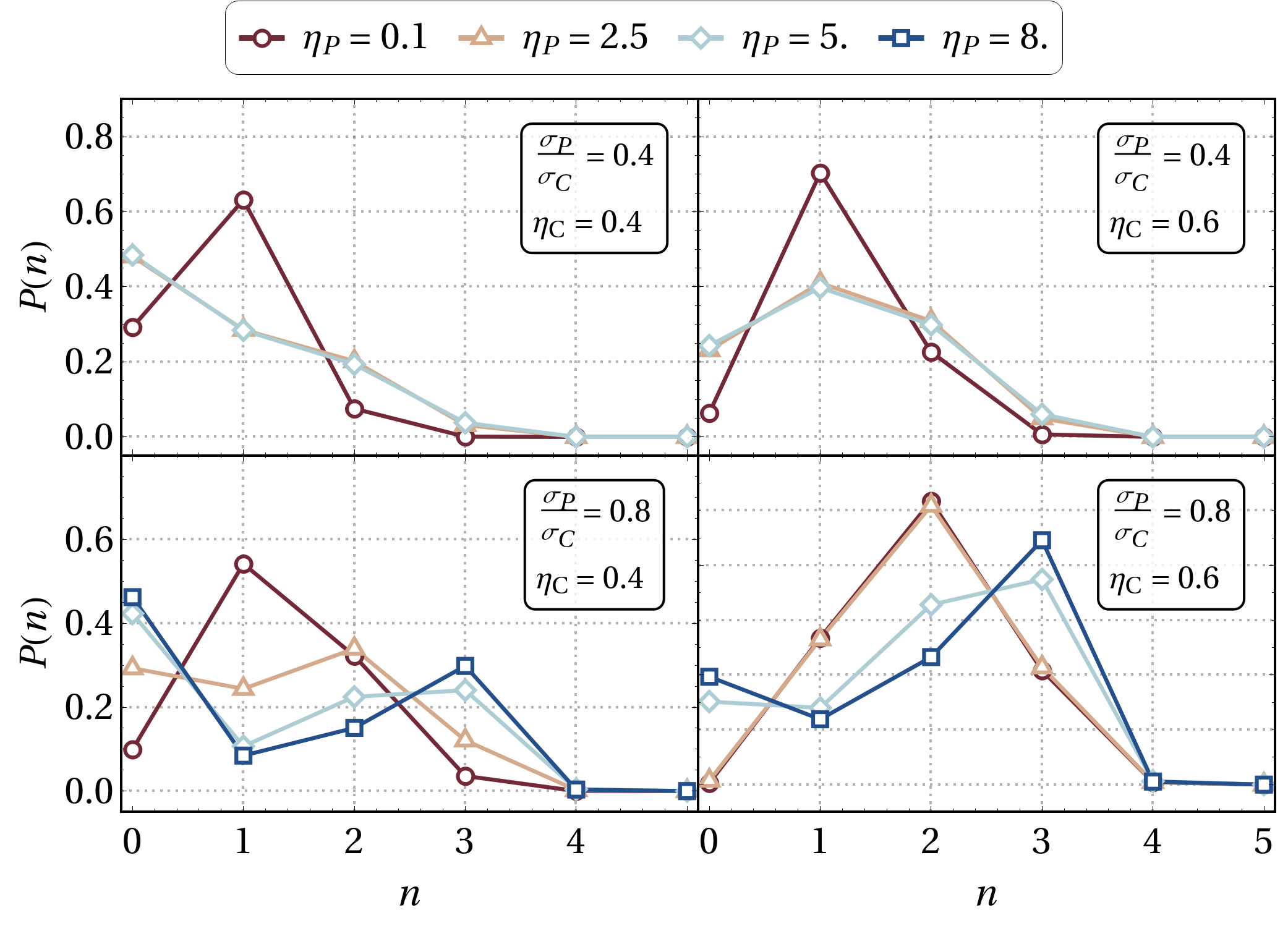}
    \caption{Probability to have an effective $n$-body colloidal interaction at an arbitrary point in space, for various polymer sizes $\sigma_P$ and various colloid and polymer packing fractions $\eta_C$ and $\eta_P$. Probability is measured using MC simulation of the  $N_C\mu_PAT$ colloid-polymer ensemble, where $N_C$ is set to 1024.}
    \label{fig:MBrealsysetm}
\end{figure}

\section{Methods}
\subsection{Model system}
We consider a 2D colloid-polymer mixture with $N_C$ colloids and $N_P$ polymers, where the total interaction potential for the system is given by 
\begin{eqnarray}
    \Phi_\text{tot} = \frac{1}{2}\sum_{i}^{N_C}\sum_{i\neq j}^{N_C} \phi_{CC}(|\mathbf{R}_i-\mathbf{R}_j|) + \sum_{i}^{N_C}\sum_{j}^{N_P} \phi_{CP}(|\mathbf{R}_i-\mathbf{r}_j|) \nonumber\\
    + \frac{1}{2}\sum_{i}^{N_P}\sum_{i\neq j}^{N_P} \phi_{PP}(|\mathbf{r}_i-\mathbf{r}_j|),
\end{eqnarray}
where, respectively, $\mathbf{R}_{i}$ and $\mathbf{r}_{i}$ denote the positions of the colloids and polymers, and $\phi_{CC}$, $\phi_{CP}$, and $\phi_{PP}$ denote the pair potentials for the colloid-colloid, colloid-polymer, and polymer-polymer interactions. In order to evaluate the performance of the potential-fitting ML algorithm, we consider a system that is governed by a set of simple interaction potentials. Specifically, we set $\phi_{PP}(r) = 0$, and choose $\phi_{CC}(r)$ and $\phi_{CP}(r)$ to be hard-disk potentials given by 
\begin{equation}
\phi^\mathrm{HS}_{ij}(r)= 
\begin{cases}
\infty &\text{for } r < \sigma_{ij}\\
0 &\text{otherwise}
\end{cases},\end{equation}
where $\sigma_{ij} = (\sigma_i + \sigma_j)/2$, with $i,j \in \{C,P\}$, and  $\sigma_C$ and $\sigma_P$ are the diameters of the colloid and the polymer, respectively. Note that this is a 2D version of the system that the Asakura-Oosawa potential is based on \cite{asakura1954interaction, vrij1976polymers}, which has been used in the past to test numerical coarse graining methods \cite{shendruk2014coarse, campos2022machine}. One of the advantages of using this simple system is that it can be readily simulated even with brute-force methods where the polymers are treated explicitly. As a result, one can easily obtain reference data for the full system. 

Our aim is to construct an effective interaction potential between the colloids, which can be used to simulate this system in a coarse-grained fashion, i.e.\ simulate only the colloids while incorporating the polymers into an effective interaction. Note that this drastically reduces the number of degrees of freedom that needs to be simulated, since in colloid-polymer mixtures the number of depletants is typically much larger than the number of colloids. In order to construct this effective potential, we turn to the ($N_C\mu_PAT$) ensemble\cite{dijkstra1999phasea}, where the number of colloids $N_C$, the area (2D volume), and the temperature are fixed, but where the number of polymers is allowed to fluctuate under the influence of a chemical potential $\mu_P$. In this ensemble, integrating out the degrees of freedom associated with the polymers leads to an effective colloid-colloid potential given by\cite{hansen2013theory}
\begin{equation}
    \Phi_\text{eff} = \frac{1}{2}\sum_{i}^{N_C}\sum_{i\neq j}^{N_C} \phi_{CC}^{HS}(\left|\mathbf{R}_i-\mathbf{R}_j\right|) +
    \Phi_\text{eff, CP}(\left\{\mathbf{R}_i\right\}),
    \label{eq:effectivepot}
\end{equation}
where
\begin{eqnarray}
\Phi_\text{eff, CP}(\left\{\mathbf{R}_i\right\}) &=& - k_B T z_P A_\text{eff}(\left\{\mathbf{R}_i\right\}) \\
A_\text{eff}(\left\{\mathbf{R}_i\right\}) &=& \int \mathrm{d}\mathbf{r}\,\exp\left[-\beta\sum_{i=1}^{N_C}\phi_{CP}(\left|\mathbf{R}_i-\mathbf{r}\right|)\right].
\label{eq:effectivepotCP}
\end{eqnarray}
Here, $z_P$ is the polymer fugacity, $k_B$ the Boltzmann constant, $\beta=1/(k_BT)$, and $A_\text{eff}(\left\{\mathbf{R}_i\right\})$ is the effective free area available to the polymers. 
This effective free area can be interpreted as a weighted area of the system, where the weight of each point in space is determined by the potential energy that a polymer at that position would experience due to the colloid configuration $\left\{\mathbf{R}_i\right\}$. Note that Eqs. \eqref{eq:effectivepot} and \eqref{eq:effectivepotCP} are valid for any arbitrary colloid-colloid and colloid-polymer interaction, provided that there are no polymer-polymer interactions.

In the case of hard-sphere colloid-polymer interactions, only points where a polymer can be placed without it overlapping with any of the colloids contribute to $A_\text{eff, CP}(\left\{\mathbf{R}_i\right\})$. As a result, $A_\text{eff, CP}(\left\{\mathbf{R}_i\right\})$ simplifies to the free area available to polymers, $A_F$.

In this paper, we focus on two polymer sizes: $\sigma_P = 0.4 \sigma_C$ and $\sigma_P = 0.8 \sigma_C$. These polymer sizes are both far outside of the regime where the approximations of the Asakura-Oosawa pair potential hold, i.e.\ where the effective colloid-colloid interaction can be written purely as a two-body interaction \cite{gast1983polymer,dijkstra1999phaseb}. Hence, for both polymer sizes, $\Phi_\text{eff, CP}(\left\{\mathbf{R}_i\right\})$ incorporates at least three-body interaction contributions. 

Note that the many-body nature of this potential arises due to the fact that more than two depletion zones (disks of radius $\sigma_{CP}/2$ around the colloid where polymer centers cannot enter) can overlap simultaneously, see Fig.\ \ref{fig:Depletion}. The probability $P(n)$ to find $n$ overlapping depletion zones at an arbitrary point in space depends on the colloid and polymer concentration, as well as the polymer size, as is shown in Fig.\ \ref{fig:MBrealsysetm}. The data in this figure is obtained by performing an MC simulation in the $N_C\mu_PAT$ ensemble, where polymers were treated explicitly. To ensure a constant $\mu_P$, we perform insertion and deletion moves of polymers via the usual Monte Carlo acceptance rule \cite{frenkel2023understanding}. To compute $P(n)$, we randomly selected points within the system and determined the number of depletion zones overlapping with these points (where points located within the colloids were included in the computation as well). For each state point, we analyzed 45 different snapshots, where we considered $10^6$ random points per snapshot. From Fig.\ \ref{fig:MBrealsysetm} we see that at lower polymer sizes there are mainly $1$- and $2$-body interactions, while at higher polymer sizes the system mainly shows $3$-body interactions.

\subsection{Capture effective potential using Voronoi cells}
Our aim is to train an ML algorithm that is able to predict the effective interactions between colloids as described in Eq.\ \eqref{eq:effectivepotCP}. However, rather than predicting the free area for the entire system at once, we aim to compute the contributions to the effective potential of individual colloids. This requires rewriting $A_\text{eff}(\left\{\mathbf{R}_i\right\})$ in terms of contributions associated with separate colloids, i.e.\ $A_\text{eff}(\left\{\mathbf{R}_i\right\}) = \sum_{i\in N_C}A^i_\text{eff}(\left\{\mathbf{R}_i\right\})$. To achieve this, each point in space must be assigned to a specific colloid, allowing the integral of Eq.\ \eqref{eq:effectivepotCP} to be split into $N_C$ integrals. A natural way to tile space is by using Voronoi cells, where each point in space is assigned to the nearest colloid, allowing us to write
\begin{equation*}
    A_\text{eff}(\left\{\mathbf{R}_i\right\}) = \sum_{i\in N_C} \int_{\mathcal{V}_i} \mathrm{d}\mathbf{r}\,\exp\left[-\beta\sum_{j}^{N_C}\phi_{CP}(\mathbf{R}_j-\mathbf{r})\right],
\end{equation*}
with $\mathcal{V}_i$ the Voronoi cell associated with particle $i$.

A key advantage of using Voronoi cells is that the shape of the cell is inherently determined by the relative positions of neighboring particles, and thus encodes information about the local many-body structure around a particle. As a result, it is possible to write the effective interactions (or in this case effective areas) into a Voronoi-based expansion: 
\begin{equation}
    A_\text{eff}(\left\{\mathbf{R}_i\right\}) = \sum_{i\in N_C} A^{(1)}({\mathcal{V}_i}) + \sum_{i,j\in N_C} A^{(2)}({\mathcal{V}_i}, {\mathcal{V}_j}) + \ldots, 
\end{equation}
where the first term in the expansion represents the contributions from single Voronoi cells, the second term takes into account corrections associated with pairs of neighbouring Voronoi cells, and so on. In the case of short-ranged interactions, the first term in the expansion will dominate, meaning that to good approximation the effective area is given by single Voronoi contributions only. In the case of hard interactions, as explored in this paper, this approximation becomes exact. The effective free area is then simply the sum of the free areas of individual Voronoi cells, i.e. 
\begin{equation*}
    A_\text{eff}(\left\{\mathbf{R}_i\right\}) = \sum_{i\in N_C} A_F^{\mathcal{V}_i},
\end{equation*}
where $A_F^{\mathcal{V}_i}$ is the free area associated with Voronoi cell $\mathcal{V}_i$. 

\begin{figure}
    \centering
    \includegraphics[width=0.5\textwidth]{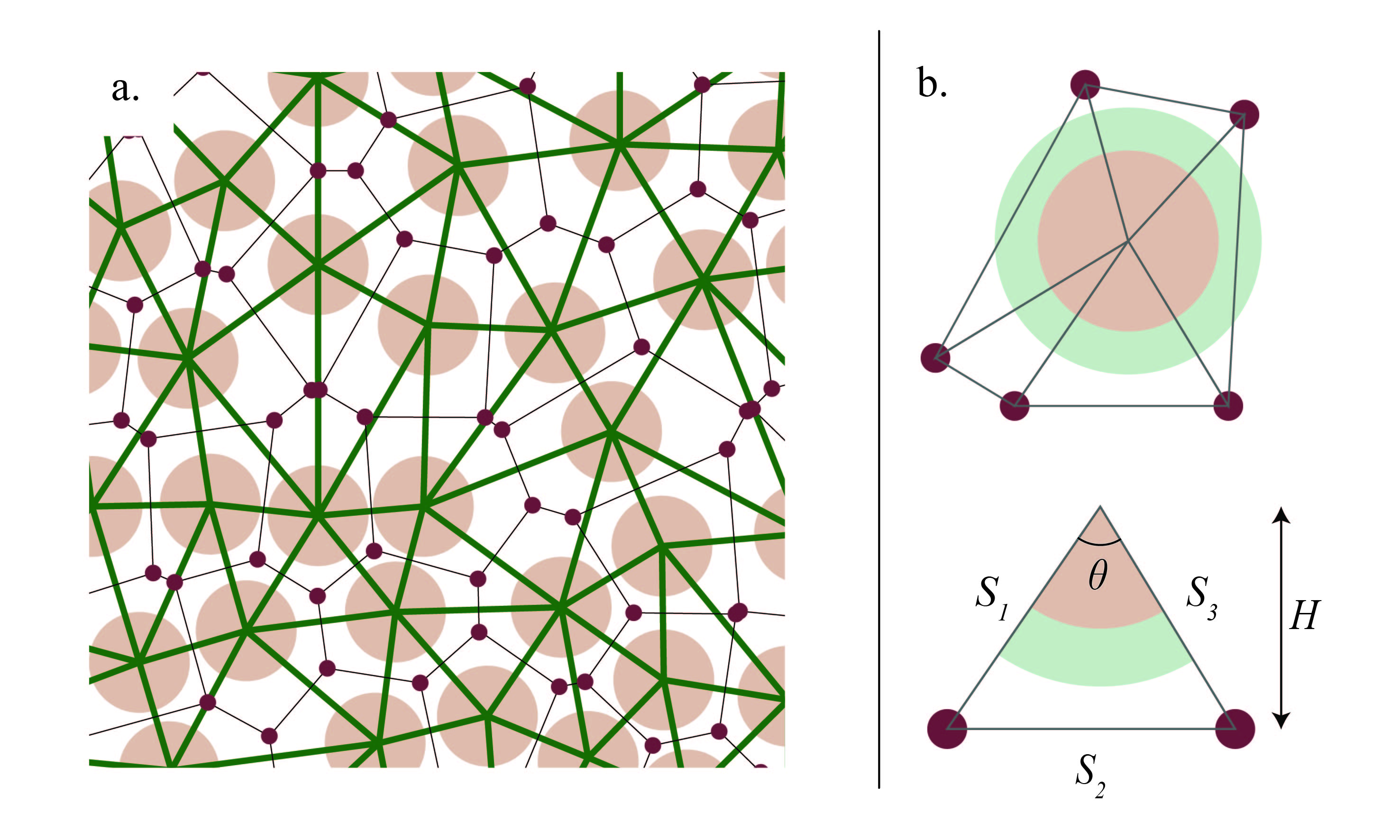}
    \caption{a. Cartoon of a  Voronoi and Delaunay tessellation for a system of colloids, where the thin, black lines represent the Voronoi tessellation and the thick, green lines represent the Delaunay tessellation. The red dots correspond both to the corners of the Voronoi cells, as well as the centers of the circumscribed circles associated with the Delaunay triangles. b. The subdivision of a Voronoi cell in terms of triangles, where each triangle connects the center of the Voronoi cell with one of the edges. The depletion zone of the Voronoi cell is shown in light green. Each triangle is represented in terms of a set of parameters as shown in the bottom figure. Additionally to the parameters shown in this figure, we also consider the perimeter $P$, the area $A_T$ and the ratio ($A_T/P^2$). Note that the free area $A_F$ of the triangle is equal to the white area inside the triangle.}
    \label{fig:Algorithm}
\end{figure}

Our goal, then, is to find a way to efficiently and in a generalizable manner, evaluate $A({\mathcal{V}_i})$, facilitating Monte Carlo simulations of the coarse-grained system. To achieve this, we use a fully-connected, feed-forward neural network to fit $A({\mathcal{V}_i})$ based on the geometry of the Voronoi cell. Therefore, we first have to characterize the Voronoi cell in terms of parameters, which can serve as input for the neural network to predict the free area. Since Voronoi cells can have a varying number of edges, it is challenging to use a fixed number of parameters to capture the cell. To address this, we subdivide each Voronoi cell into triangles, where each triangle connects one of the edges to the center of the cell (see Fig.\ \ref{fig:Algorithm}). We can then apply the neural-network fit to each triangle individually, after which the effective potential of the entire Voronoi cell is obtained by summing over the outputs of all triangles. Note that for this system, this approach is exact, but it would become an approximation for interacting polymers. 

To capture the triangles in terms of parameters, we use seven different parameters (see Fig.\ \ref{fig:Algorithm}b): the area $A_T$, the lengths of the three sides, $S_1$, $S_2$ and $S_3$, the perimeter $P$, the angle $\theta$ between the two sides that come together in the center of the Voronoi cell, the height $H$ of the triangle with respect to the side that is not connected to the particle, and finally the ratio $R$ between the area and the perimeter squared ($A_T/P^2$). Note that these parameters were not optimized, but rather found to be sufficient to capture the geometry of the triangles.

\subsection{Fitting the free area of a Voronoi cell}
The subsequent step is to design an ML algorithm that predicts the free area of a triangle given its input parameters. Since the aim is to implement the algorithm in an MC simulation, its complexity is relatively constrained to ensure fast simulations. Here, we experimented with two small fully connected neural networks with respectively 3 ($[3, 3, 3]$) and 4 ($[5, 5, 3, 3]$) hidden layers. As an activation function, we used a Rectified Linear Unit (ReLU) and the framework employed to train the model is the Python package PyTorch\cite{PYTORCH}, together with an Adam optimizer\cite{adam}. We used a batch size of $100$, a learning rate of $0.0001$ and $250$ epochs. For the loss function, we experimented with two different functions. First, we considered the mean square error (MSE), a common loss function to train ML algorithms with, defined as
\begin{eqnarray}
 L_\text{MSE} &=& \frac{1}{N}\sum^N(y_i-\bar{y}_i)^2\\
\label{eq:lossMSE}
\end{eqnarray}
with $N$ the number of data points, $y_i$ the predicted value and $\bar{y}_i$ the true value. The second loss function we considered, is the mean square logarithmic error (MSLE), which is given by 
\begin{equation}
L_\text{MSLE}=\frac{1}{N}\sum^N[\log(a+y_i)-\log(a+\bar{y}_i)]^2,
\label{eq:losslog}
\end{equation}
with $a$ a constant that we choose equal to $\sigma_C^2$. With this choice of $a$, combined with the logarithmic nature of the MSLE, the loss function puts less weight on the errors associated with large values. As such, the function is a natural choice for systems, like the depletant system, where the range of possible values is large.

\subsubsection{Training data}
\begin{figure}
    \centering
    \includegraphics[width=0.45\textwidth]{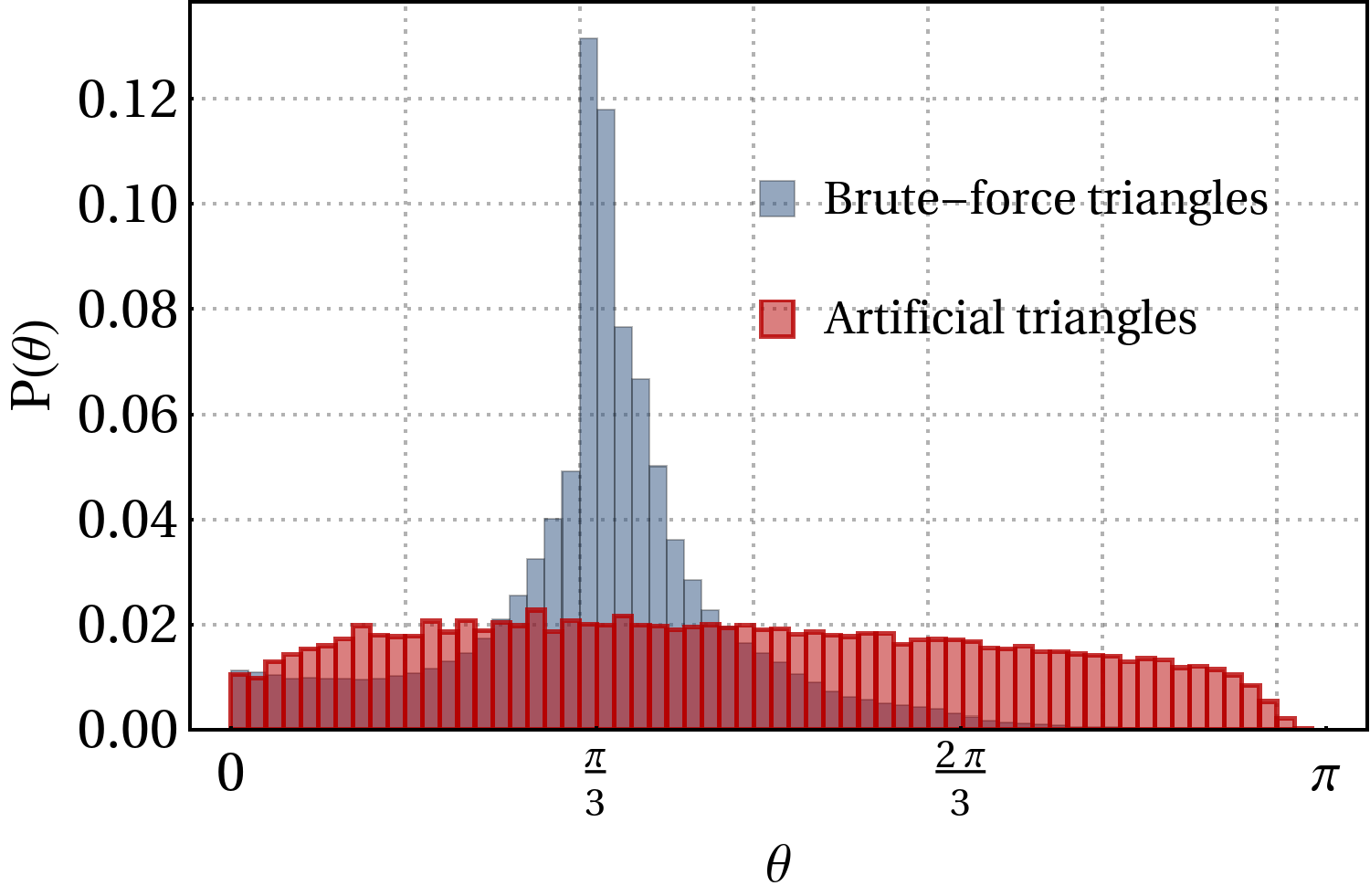}
    \caption{Probability $P(\theta)$ to have a certain angle $\theta$ in the training dataset for a polymer size $\sigma_P = 0.8 \sigma_C$, where  $\theta$ is the angle between the two sides that come together in the center of the Voronoi cell. Blue data represent the distribution as found in the brute-force training data set, while the red data represent the distribution obtained from the artificially generated training data set. Distributions are based on approximately $6\cdot10^5$ datapoints.}
    \label{fig:histogramangle}
\end{figure}
A common method to build a training dataset for machine learning potentials is to use configurations of the full system as input for the ML training. However, as has been seen before, this approach can lead to an unbalanced training data set when certain environments are under- or overrepresented in the training data. In this specific case, we found that configurations from the brute-force simulation contained an abundance of small equilateral triangles, stemming from the fact that at higher polymer densities the depletion system phase separates into a polymer gas and a dense 2D colloidal hexagonal crystal. To demonstrate this, in Fig.\ \ref{fig:histogramangle}, we show the probability $P(\theta)$ to encounter a triangle with an angle $\theta$ in the brute-force training data. Here, the training data is obtained via brute-force simulations, for a polymer size $\sigma_P = 0.8 \sigma_C$ at colloid packing fractions $\eta_C$ between $0.05$ and $0.9$ and average polymer packing fractions $\langle\eta_P\rangle\in{[0.005, 2.5,5, 8]}$, where $\langle\eta_p\rangle = \frac{\pi\sigma^2_P}{4}e^{\beta\mu_P}$. Due to the abundance of triangles with an angle around $\pi/3$ in the training data, we found that models trained on this data did not generalize well. 

While in general various strategies have been devised to circumvent the issue of an unbalanced training data set (e.g. by iteratively enhancing the training set) \cite{jinnouchi2019phase}, the simple nature of this model opened the door for another approach. Instead of using configurations of the full system, we opted to generate artificial training points. In general, one could generate Voronoi cells, or even entire artificial configurations. However, since for the considered system each point in the training data consists of a triangle (with its relevant shape parameters), and the associated free area, we can generate training data simply by creating random triangles and calculating the associated free area. To generate the triangles which make up the Voronoi cells, we use the following scheme. 

\begin{enumerate}
    \item We create a triangle by generating two random numbers between $0$ and $1$ that correspond to the sides connected to the particle ($S_1$ and $S_3$) and a random angle in the range $\theta \in [0, \pi]$ between those two sides. This generates triangles with an essentially arbitrary shape.
    \item We rescale the triangle such that either $S_1$, $S_3$ or the height of the triangle $H$, is equal to $\sigma_C/2$ while the other two parameters are larger than $\sigma_C/2$. Note that triangles that do not satisfy this would inevitably lead to overlaps between the associated colloids.
    \item We rescale the triangles twenty times to different sizes of the same shape. Specifically, we scale the area of each triangle 20 times (starting from the same initial size), 10 times by a random factor $a$ with $a\in[1, 4\sigma_{cp}^2/\sigma_{c}^2]$ and 10 times by a random factor $a\in[4\sigma_{cp}^2/\sigma_{c}^2, 1000]$. From these triangles we only use the triangles for which it is true that $S_1, S_3, H < 20\sigma_C$. 
    \item For all the accepted triangles, we compute the free area, $A_F$. For certain triangles, the free area can be easily calculated analytically, namely
    \begin{equation}
    A_F =
    \begin{cases}
    0 &\text{ if } S_1, S_3 < \sigma_{CP}\\\
    A_T -\frac{\theta}{2\pi} \pi\frac{\sigma_{CP}^2}{4} &\text{ if } S_1, S_3, H > \sigma_{CP}
    \end{cases}.
    \label{eq:trivtriangle}
    \end{equation}
    If the free area is non-trivial, we determine it numerically by Monte Carlo integration. 
\end{enumerate}
As shown in Eq. \eqref{eq:trivtriangle}, there is a subset of triangles for which the free area is trivial. In the MC simulation of the coarse grained system, we analytically calculate the free area of those triangles, and thus only use the neural network to predict the free area for the non-trivial cases. Note, however, that we still included the trivial triangles in our training data because it ensures that the extreme values of the triangles for which the NN has to be used, lie well within the boundaries of the training dataset. As a result, the NN has to extrapolate less at the boundaries. Moreover, in the explicit simulations we saw many triangles with a small $\theta$. Therefore, we ensured that the lower tail of the $\theta$ distribution was well represented in the training data; for every $20,000$ triangles we always included $100$ triangles with a random $\theta < 0.0001$ and  $100$ triangles with a random $\theta < 0.1$. Since we observed that triangles with $\theta \approx \pi$ almost never occur in the actual system, we did not include additional triangles with a large $\theta$ in the training data.
To compare the artificially generated training data with the brute-force training data, in Fig.\ \ref{fig:histogramangle} we also show $P(\theta)$ for the artificially generated triangles. As we can see in this figure, the artificially generated triangles lead to a significantly better balanced training data set.

\begin{figure}
    \centering
    \includegraphics[width=0.4\textwidth]{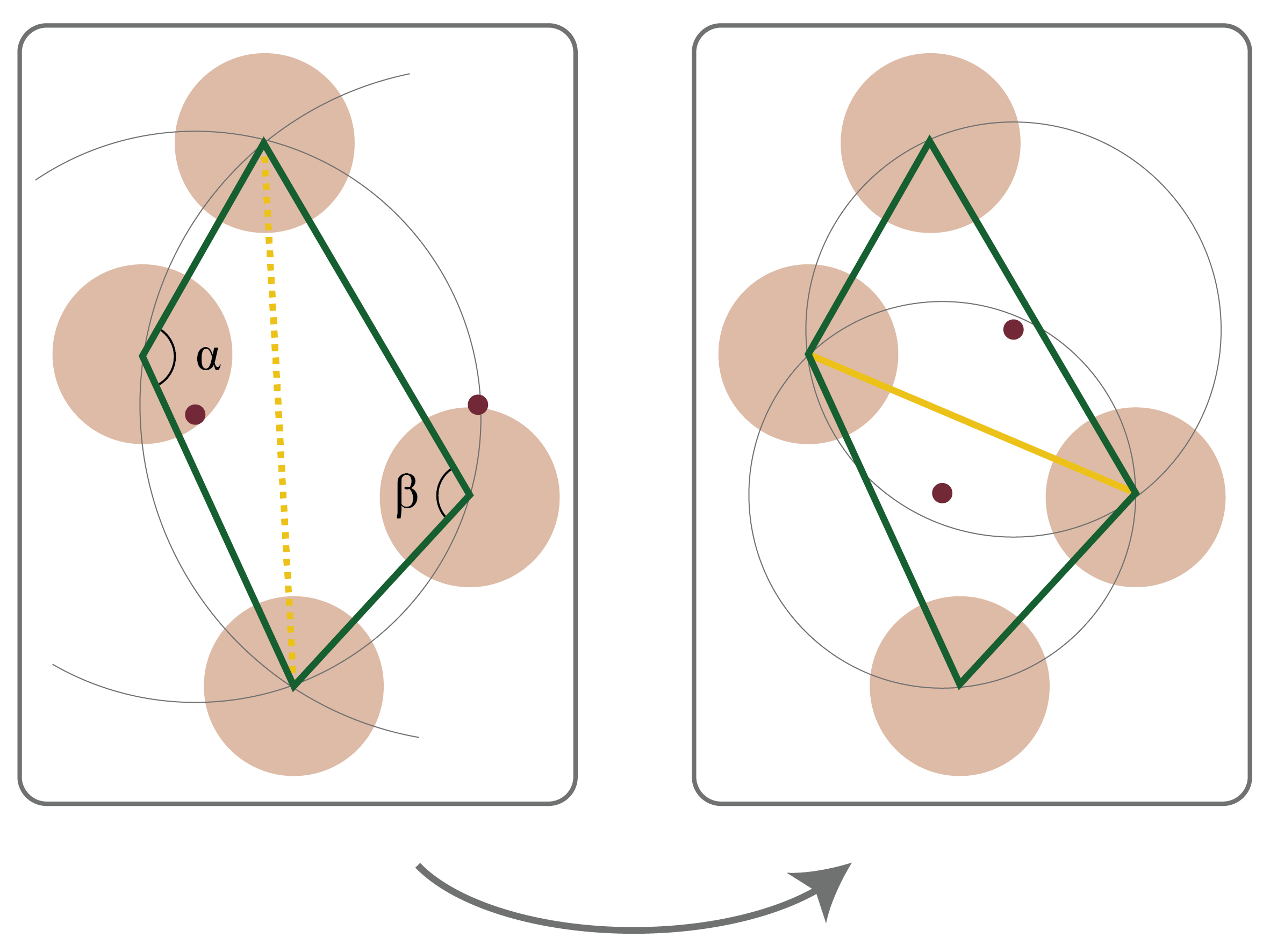}
    \caption{Cartoon of a flip move in the equiangulation procedure that it is used to obtain a correct Delaunay tessellation. If two adjacent Delaunay triangles share opposite angles that together are larger than $\pi$, i.e. $\angle \alpha +\angle \beta > \pi$, the edge that the two triangles share is flipped. }
    \label{fig:flipmove}
\end{figure}

\subsection{Implement algorithm in Monte Carlo simulation}
\label{sec:MCimplementation}
In order to be able to test the quality of the ML potential, we have to implement the Voronoi tessellation in an MC  simulation. In these MC simulations, trial particle moves are accepted or declined according to the overall Boltzmann weight of the system.  For the system under consideration, this Boltzmann weight solely depends on the change in free area of the Voronoi cell of the displaced particle and its neighbors, before and after the move. Simulations of Voronoi tesselations have been widely used in the context of vertex models to simulate e.g.\ tissue, and in this paper we build on the developments in this field \cite{honda1980much,bi2016motility,barton2017active}. 
It is numerically very costly to construct a new Voronoi tessellation (VT) from scratch for each new configuration\cite{barton2017active}. However, since the difference between consecutive configurations and thus tessellations in an MC simulation will be minimal, we can instead use the old tessellation as a starting point to obtain the new tessellation\cite{barton2017active}. Although we are not aware of an algorithm that directly updates a VT, we can circumvent this problem by turning to the dual of the Voronoi graph, namely the Delaunay tessellation (DT). In the DT, every two particles that share an edge in the VT are connected, such that the DT divides the area into triangles that always connect three particles (see Fig.\ \ref{fig:Algorithm}a). Given an arbitrary tessellation, one can converge to the actual DT by applying the equiangulation procedure \cite{brakke1992surface}. This procedure makes use of the fact that in a correct DT the circumscribed circle of each of the triangles cannot contain any other vertices. If two adjacent triangles have opposite angles that are together larger than $\pi$, this circumscribed criterion is no longer met, meaning that the shared edge should flip (see Fig.\ \ref{fig:flipmove}). By iteratively switching edges, it can be shown that the system always converges to the correct DT\cite{brakke1992surface}. Given this correct DT, one can then convert back to the VT and compute the free area associated with that configuration. In the SI we discuss in more detail how we realized the implementation of the Delaunay tessellation in an MC simulation.

In our simulations, we always start from a correct DT based on a configuration of colloids in a gas configuration. This initial DT can easily be obtained by using one of the existing Delaunay triangulation packages. In this work, we used Ref.\ \onlinecite{2020SciPy-NMeth}.

\begin{figure*}
    \centering
    \begin{tabular}{lc}
     a) &  \\[0cm]
     & \includegraphics[width=0.8\linewidth]{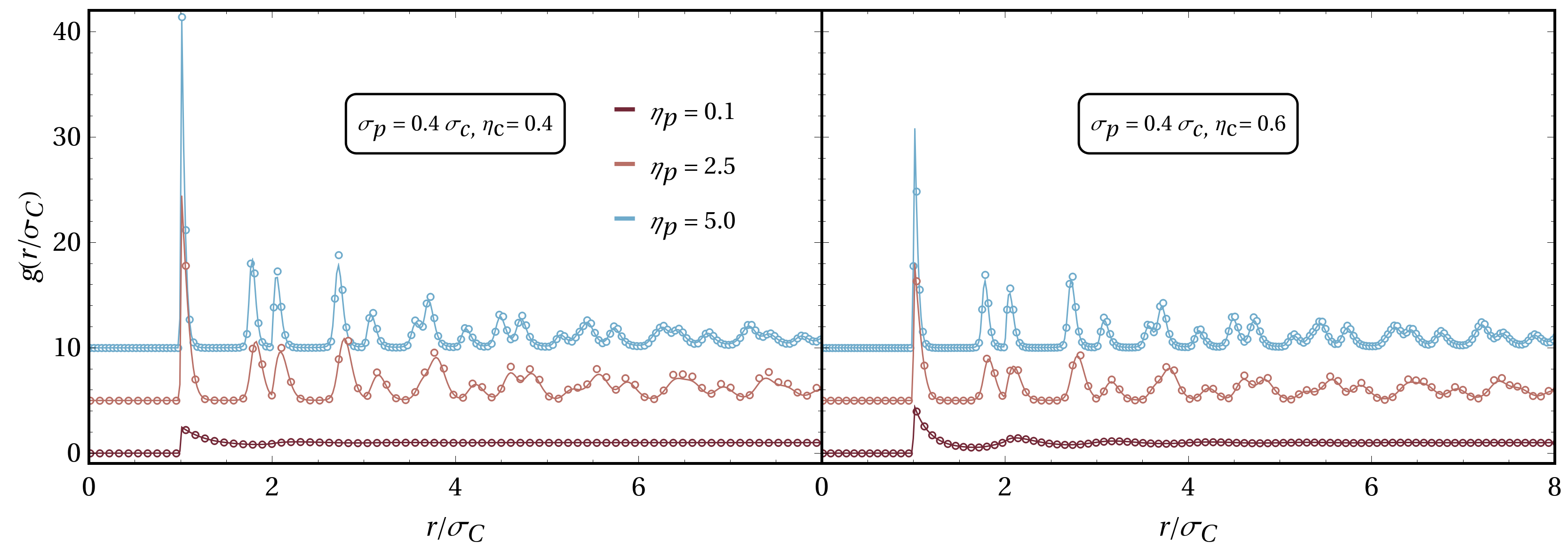}\\   
      b) &  \\[0cm]
       & \includegraphics[width=0.8\linewidth]{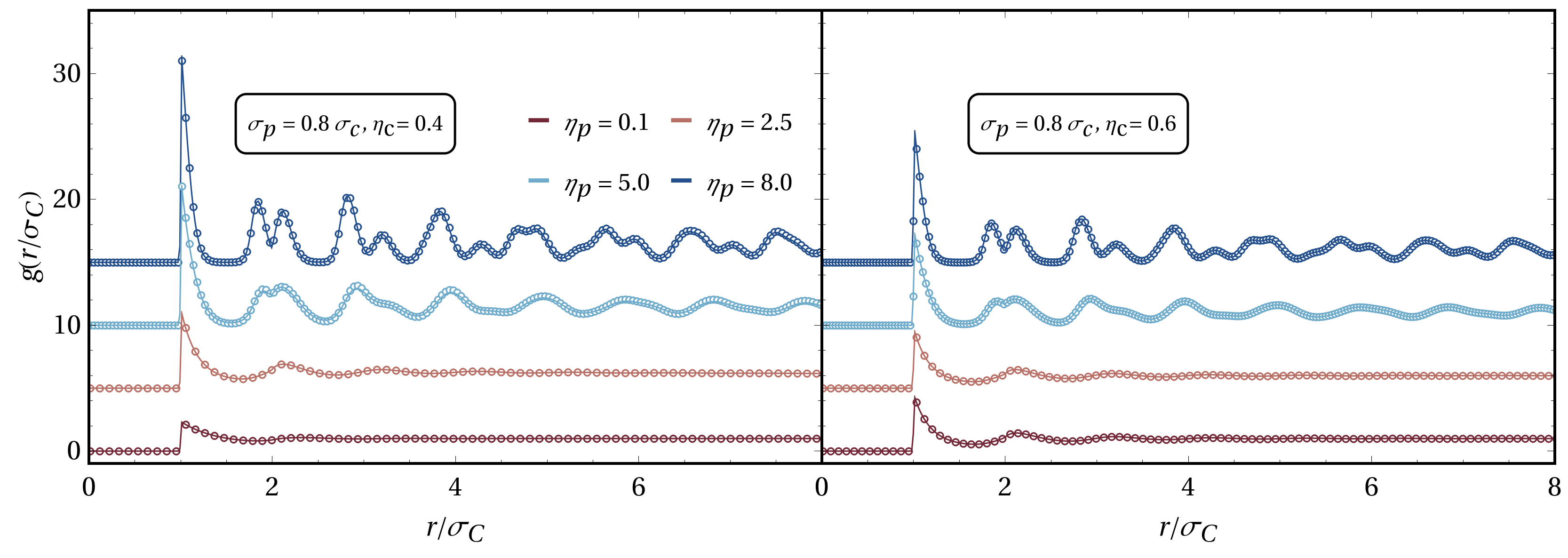}\\ 
\end{tabular}
    \caption{The pair distribution function $g(r/\sigma_C)$ plotted for a polymer size of $\sigma_P = 0.4$ and $\sigma_P = 0.8$ in respectively panel a) and panel b) for various polymer packing fractions $\eta_P\in[0.1, 2.5,5.0, 8.0]$ and two colloid packing fractions $\eta_C\in[0.4, 0.6]$. The solid lines represent the ground truth as measured in the system where polymers are treated explicitly whereas the plot markers represent the results as measured in the ML systems. Both the solid lines as well as the plot markers are averaged over five different simulations. For clarity, the radial distribution functions are shifted vertically, with higher curves corresponding to systems with larger polymer fractions.}
    \label{fig:resultspaircor}
\end{figure*}

\begin{figure*}
    \centering
    \includegraphics[width=0.7\textwidth]{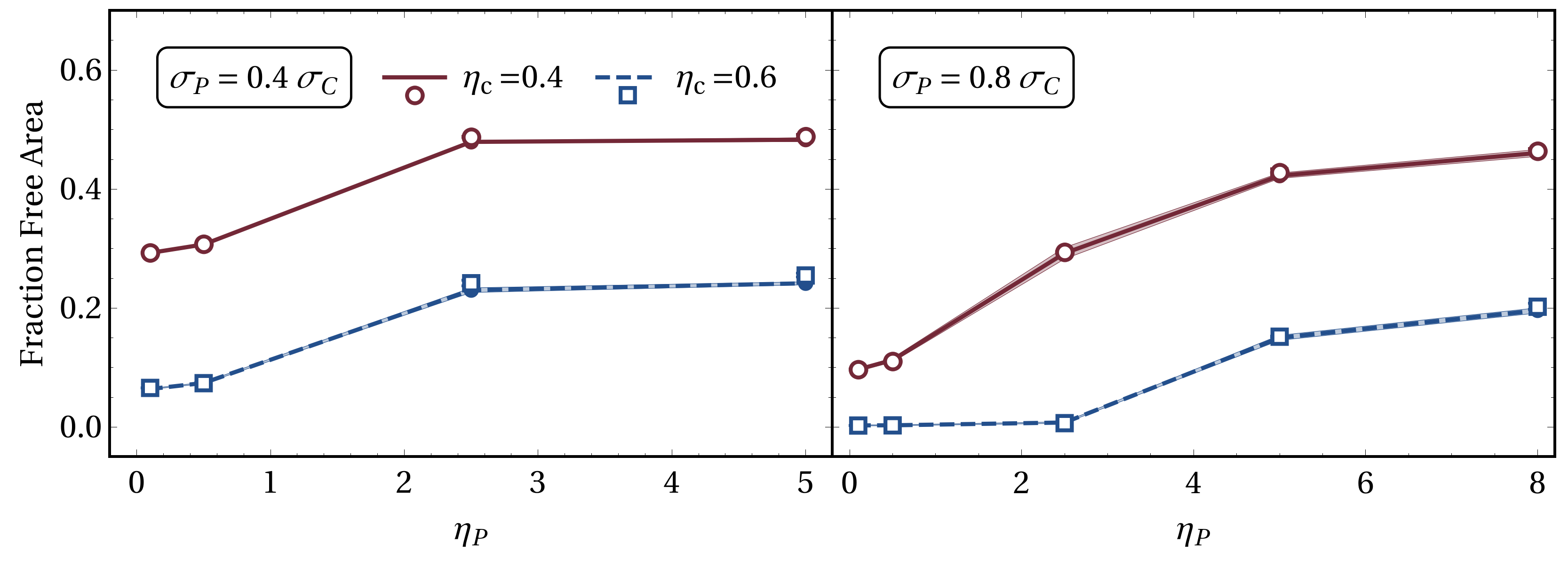}
    \caption{The fraction of free area plotted as a function of the polymer packing fraction $\eta_P$ for two polymer sizes, $\sigma_P = 0.4 \sigma_C$ (left) and $\sigma_P = 0.8 \sigma_C$ (right) and two colloid packing fractions $\eta_C=[0.4, 0.6]$. The lines represent the ground truth as measured in the system where polymers are treated explicitly (here, the small transparent region indicates the standard deviation) The plot markers represent the free area fraction measured in the system simulated with the ML potential.}
    \label{fig:freevol0408}
\end{figure*}

\begin{figure*}
    \centering
    \includegraphics[width=0.7\textwidth]{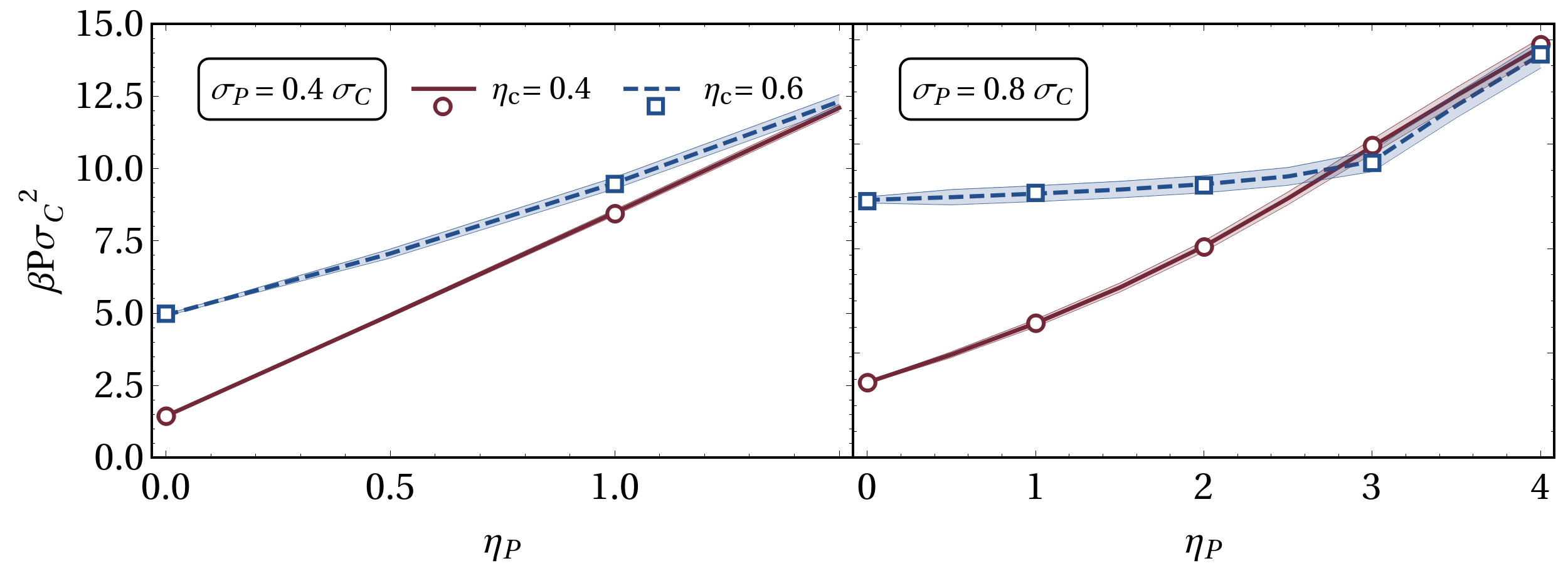}
    \caption{The pressure plotted as a function of the polymer packing fraction $\eta_P$ for two polymer sizes, $\sigma_P = 0.4 \sigma_C$ (left) and $\sigma_P = 0.8 \sigma_C$ (right) and two colloid packing fractions $\eta_C=[0.4, 0.6]$. The lines represent the ground truth as measured in the system where polymers are treated explicitly (here, the small transparent region indicates the standard deviation). The plot markers represent the pressure measured in the system simulated with the machine learned potential.}
    \label{fig:press0408}
\end{figure*}

\subsection{Training and simulation details}
In order to train the ML algorithms, we used a dataset containing $10^7$ triangles. Half of this data was used as training data and half was used as test data. We trained the model separately for two polymer sizes $\sigma_P\in[0.4, 0.8]$ and different sets of input parameters.

The ML simulations are implemented in the $N_CVT$ ensemble, and contain $1024$ colloids. Each simulation is equilibrated for $2\cdot 10^7$ MC cycles, after which measurements are collected for $10^7$ MC cycles.

To test the result of the trained model, we compare the ML simulations to brute-force simulations. These brute-force simulations are obtained using both MC and Event Driven Molecular Dynamics (EDMD)\cite{rapaport2009event} simulations, both implemented in the earlier mentioned $N_C\mu_PAT$ ensemble, where  $N_C$ is set to $1024$. The MC simulations are equilibrated for $10^7$ MC steps, after which measurements are collected for $10^7$ MC steps. The EDMD simulations were run for $t/\tau = 10^5$ with $\tau = \sqrt{\frac{m\sigma^2}{k_BT}}$, where one-fourth of this time was used as equilibration time. For the EDMD simulations we adapted the code of Ref. \onlinecite{smallenburg2022efficient}.  

In order to validate the accuracy of the ML potentials, we studied how well the ML potential reproduces the structure of the system by comparing the radial distribution function of the colloid-colloid interactions. Additionally, we measured the free area fractions ($A_\mathrm{eff}/A$) to examine whether the potential energies match. Both of these measurements were tested against the brute-force simulations obtained via MC simulations. 

Finally, we compared the virial pressures between the ML system and the brute-force system. The pressure measurement of the brute-force system is performed during an EDMD simulation by tracking the momentum transfer associated with each collision\cite{alder1960studies}. In order to measure the pressure in the ML system, we adopted a method analogous to the Widom test insertion method\cite{widom1963some}. Starting from the thermodynamic relation $ P= -\frac{\partial F(N, A, T)}{\partial A}$, we approximate it as 
\begin{align*} 
    P =-\frac{1}{\Delta A}\left[F(N, A+\Delta A, T)- F(N, A, T)\right],
\end{align*} 
where $F(N, A, T)$ is the free-energy for a system with fixed $N$, $A$ and $T$. By expressing the free energy in terms of the partition function, the free energy difference can be written as
\begin{align*}
F(N,&\, A+\Delta A, T)- F(N, A, T) \\&=-k_BT\log\left(\frac{Z_{A+\Delta A}}{Z_A}\right)\\
    &=-k_BT\log\left\langle e^{-\beta\left[ U(\mathbf{s}^N, A+ \Delta A) - U(\mathbf{s}^N, A)-Nk_BT\log\frac{A + \Delta A}{A}\right]}\right\rangle,
\end{align*}
where $Z_A$ is the partition function of a system with area $A$, where $\mathbf{s}^N$ are the scaled colloid coordinates, and where $\langle\dots\rangle$ denotes an ensemble average. Thus, the pressure can be measured during an MC simulation by performing a series of (small) trial area changes and averaging the corresponding potential energy difference. Note that since scaling the system does not affect the Voronoi tesselation, performing a pressure measurement in the ML simulation is computationally cheap.
\section{Results}

\begin{figure*}[t!]
\begin{tabular}{lclc}
     a) &  & b) &   \\[0cm]
     &  \includegraphics[width=0.45\linewidth]{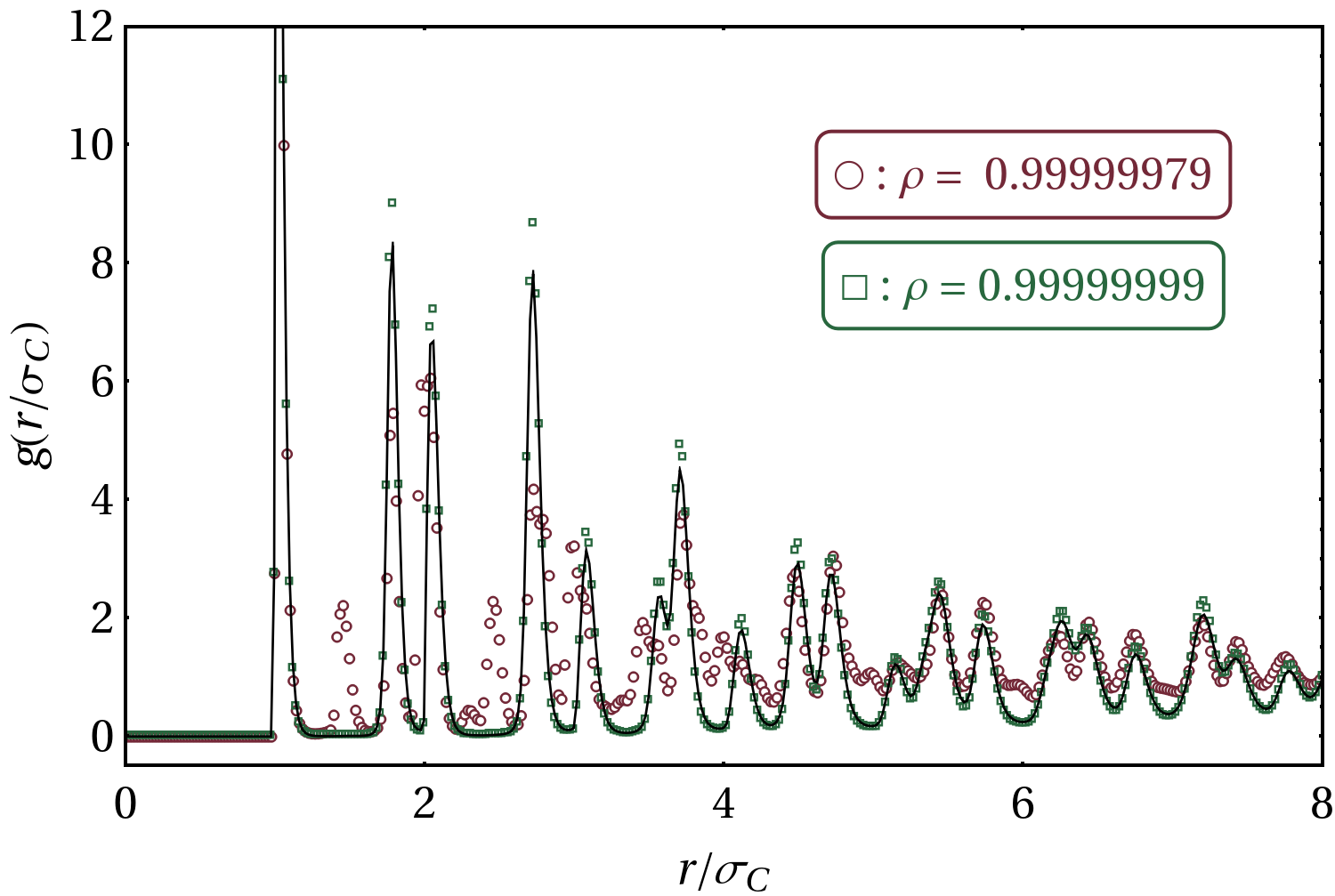} & &\includegraphics[width=0.45\linewidth]{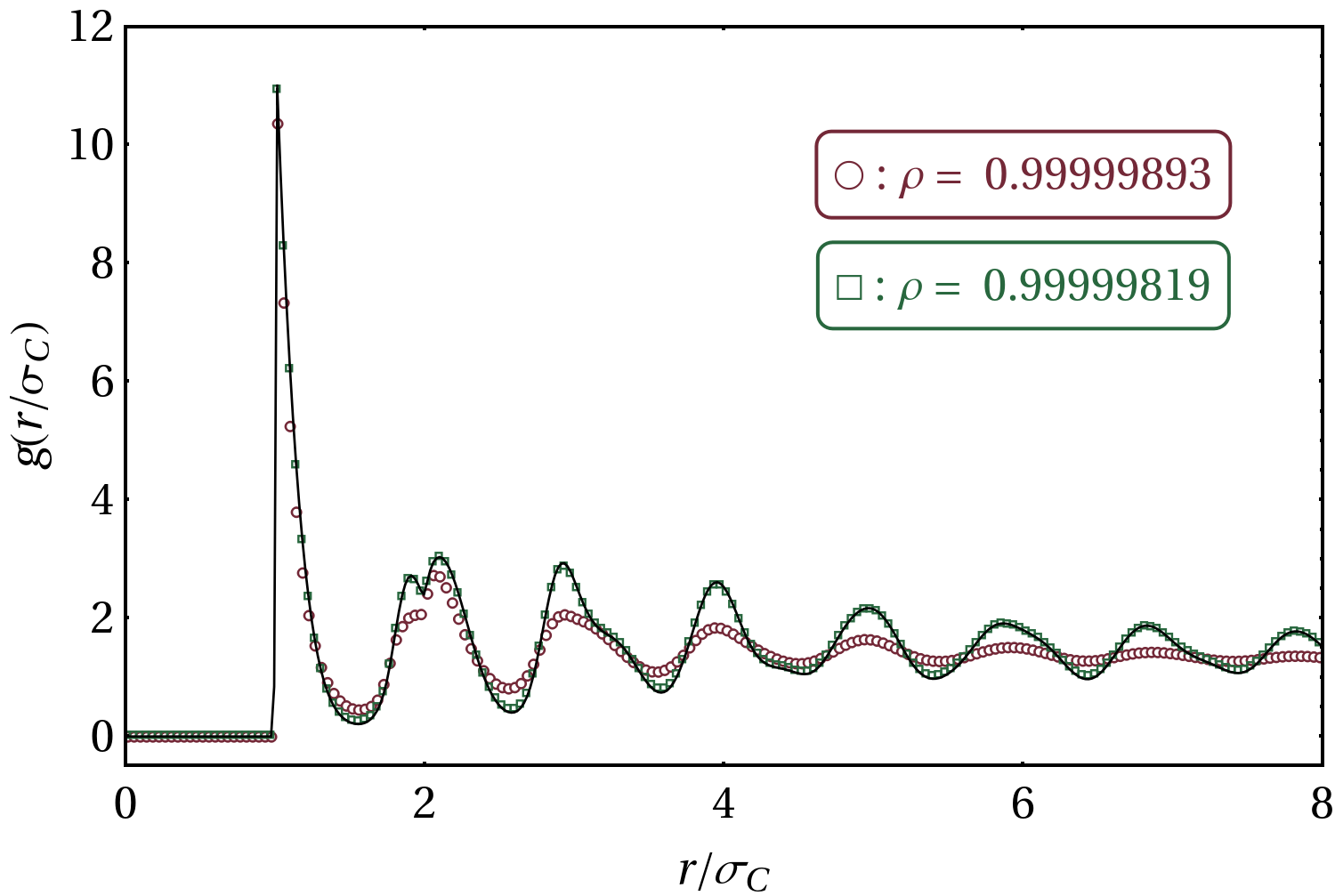} \\
        c) &  & d) &   \\[0cm]
     &  \includegraphics[width=0.45\linewidth]{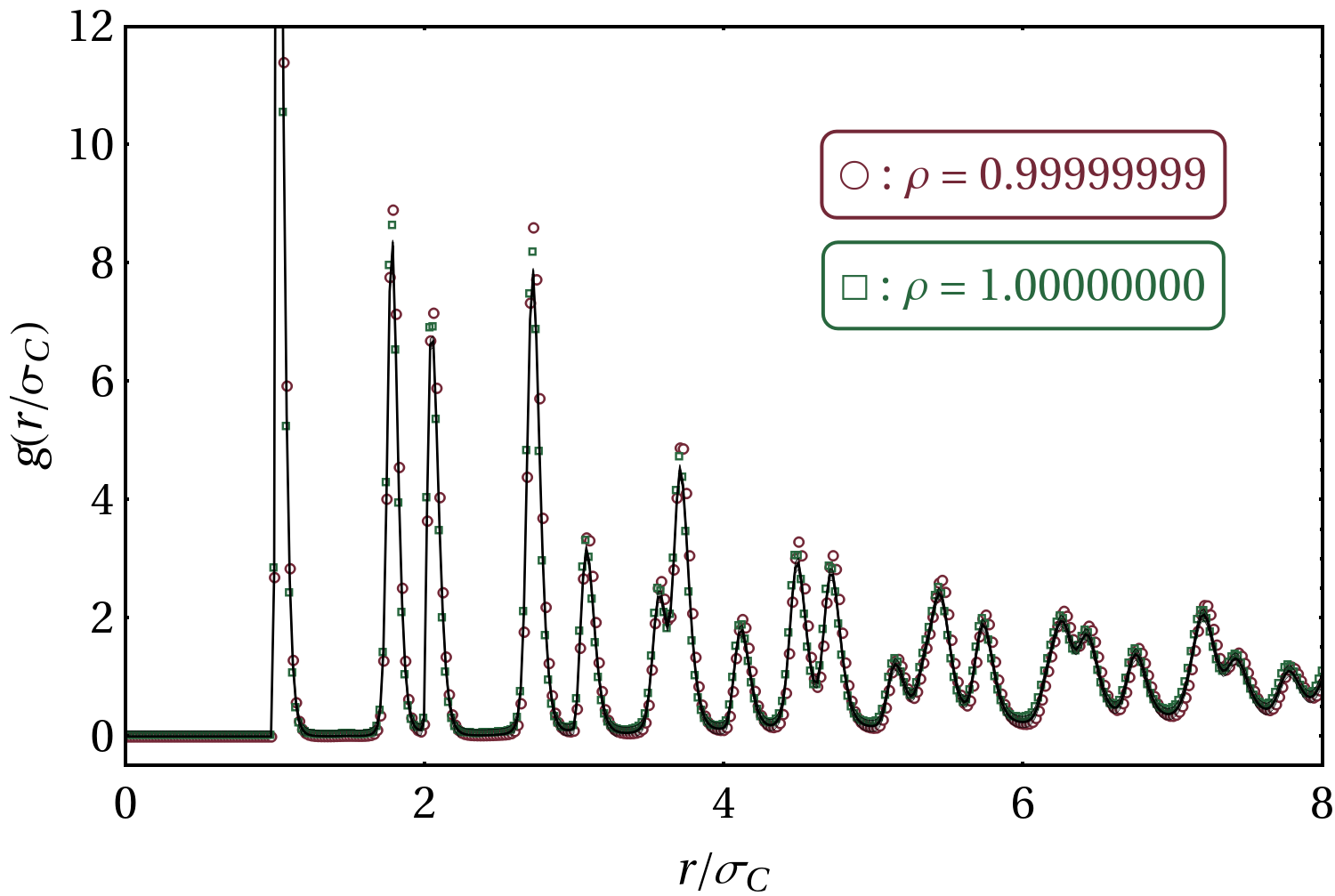} & &\includegraphics[width=0.45\linewidth]{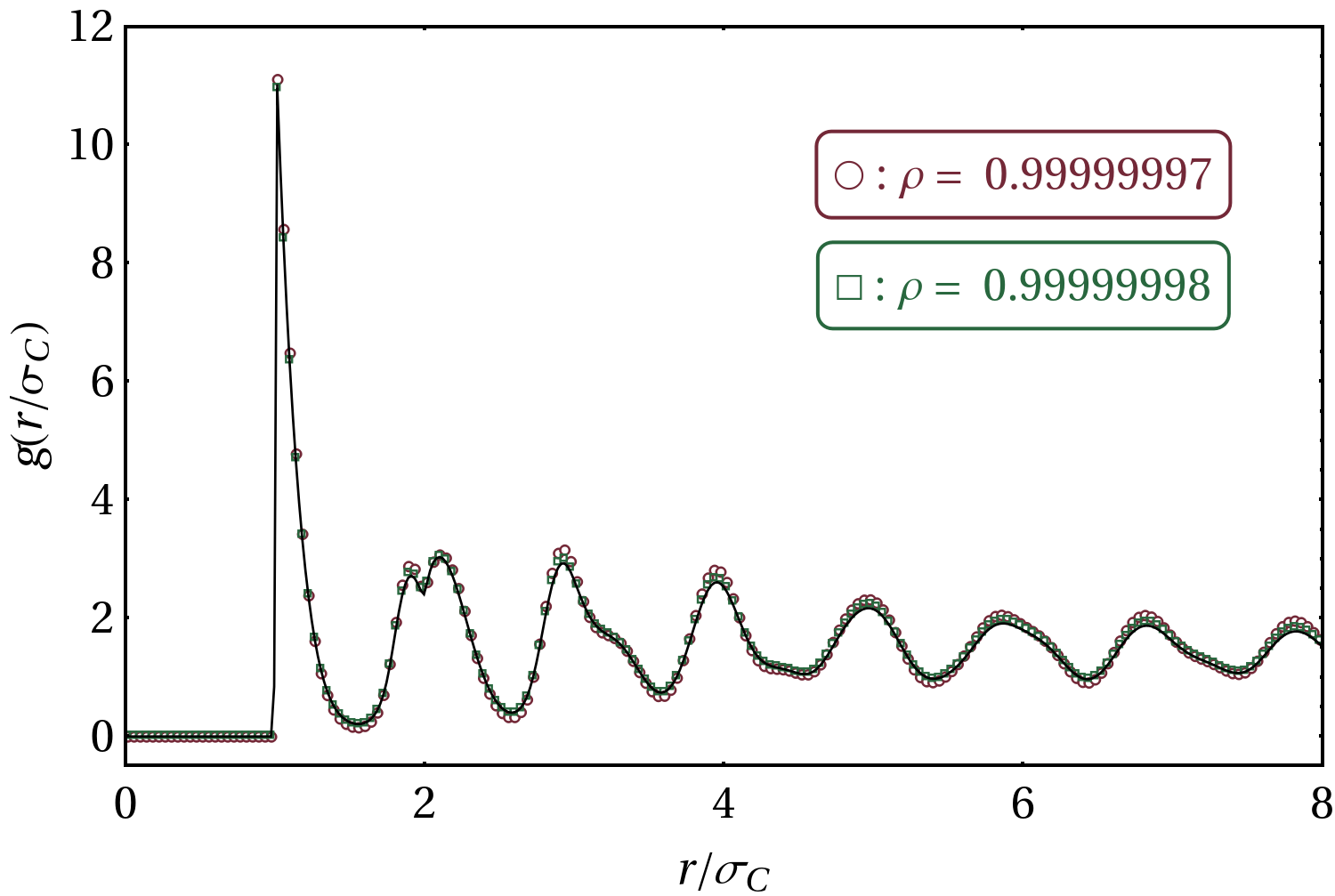} 
\end{tabular}
    \caption[width=1\linewidth]{Best (green square markers) and worst (red circle markers) radial distribution functions, selected by eye, from six independently models that were trained on all input parameters.  In panel a) and c), results are obtained using a model that was trained with the MSLE loss function on a polymer size $\sigma_P = 0.4$ with respectively 3 hidden layer NN with respectively $[3, 3, 3]$ nodes and a 5 hidden layer NN with respectively $[5, 5, 3, 3]$ nodes. In panel b) and d), results are obtained using a model that was trained with an MSE loss function on a polymer size $\sigma_P = 0.8$ the small and large NN size respectively. All systems have a polymer packing fraction of $\eta_P=5.0$ and a colloid packing fraction of $\eta_C=0.4$. In the plots, the solid lines represent the ground truth as measured in the system where polymers are treated explicitly. Note that in the plot the Pearson correlation $\rho$ between true and predicted free area values is indicated for both runs.}
    \label{fig:comparisonnetwork} 
\end{figure*}
Our aim is to train an ML model that is able to consistently and accurately reproduce the correct structure in a depletion system given a certain polymer size. As mentioned earlier, we test our methodology by comparing the radial distribution functions between the ML model and the full system. Note that reproducing the correct radial distribution function implies that many of the thermodynamic properties are correct \cite{hansen2013theory}. We also show this thermodynamic consistency explicitly by comparing the virial pressure between the brute-force system and the ML system. 

We begin by looking at the results of a NN with 4 hidden layers consisting of ($[5, 5, 3, 3]$) nodes which was trained with the MSLE loss function.  Note that we train a single NN per polymer size.
In Fig.\ \ref{fig:resultspaircor}a) and b) we plot the colloid-colloid radial distribution functions for $\sigma_P = 0.4\sigma_C$ and $\sigma_P = 0.8\sigma_C$ respectively, for a collection $\eta_C$'s and $\eta_P$'s. In this figure, the solid lines represent the radial distribution functions as measured in the full system, where each line is averaged over five independent simulations. The plot markers correspond to the radial distribution function as measured in the ML simulations, where we again averaged over five independent simulations. The excellent agreement between the two models clearly shows that the ML potential is able to accurately reproduce the many-body potential, for a wide range of polymer and colloidal densities. 

In order to test whether this combination of network size and loss function leads to models that perform consistently, we trained six independent models with identical hyperparameters for both polymer sizes. As shown in the appendix (right panels of Figs.\ \ref{fig:resultspaircor04bigNNloss} and \ref{fig:resultspaircor08bigNNloss}), this combination of hyperparameters leads to models that are able to consistently reproduce the correct radial distribution function. 

To further test the accuracy of our models, for a single model we measured the free area fraction $A_\mathrm{eff}/A$  as a function of the polymer packing fraction for the same phase points as displayed in Fig.\ \ref{fig:resultspaircor}. The results are shown in Fig.\ \ref{fig:freevol0408}, where the lines represent the full system, the plot markers represent the ML-systems, and where in both cases the results are averaged over five independent simulations. Again, we observe excellent agreement between the full system and the ML potential.  

As an additional test of our model, we compare the virial pressures of the ML model and the brute-force model. In Fig.\ \ref{fig:press0408}, we show the pressure for two different polymer sizes and a range of different colloid and polymer packing fractions (note that we only include polymer and colloid packing fractions that are below colloidal crystallization).  
As we can see, the two pressure measurements excellently agree with each other.

At this point, it is interesting to examine the relative simulation speed obtained from the ML-driven code in comparison to our full-model simulations. Depending on the polymer size and packing fraction, we observe a significant speed up in simulation time between the ML-system and the brute-force system MC system. For example, for $\sigma_P=0.4$  and polymer packing fractions above $2.5$, the ML simulations were roughly $10$ to $50$ times faster. This speed-up is all the more impressive when considering the fact that the system we consider here was chosen to be fast to simulate brute-force, while simultaneously having non-trivial many-body contributions.  

From these results, we can conclude that we achieved our goal of training an ML model based on the Voronoi structure, that is able to consistently and accurately reproduce the correct structure.  However, during the training process, we made some important observations which we outline here. 

\textit{Loss Functions:} As outlined in the methods, we explored the influence of two different loss functions on the accuracy of the models. Between the loss functions that we tested (MSE and MSLE), we found no significant differences (as shown in Appendix Figs. \ref{fig:resultspaircor04bigNNloss} and \ref{fig:resultspaircor08bigNNloss}).  

\textit{Network Size:} In contrast to the loss functions, we found that the network size did have an influence on the performance of the models. Although the smaller network size of 3 hidden layers with ($[3, 3, 3]$) nodes was able to reproduce the correct structure most of the time, we found that its performance was less consistent compared to the larger network size. To explore this, we again trained 6 different models for the smaller network and compared the results to that of the larger network.  In Fig. \ref{fig:comparisonnetwork} we show the best (green) and worst (red) radial distributions from the six trained models (identified by eye). Here panel a) shows the results for models trained on $\sigma_C=0.4\sigma_C$ with the MSLE for the phase point $\eta_P=5.0$, $\eta_C=0.4$, while panel b) shows the results for models trained on $\sigma_C=0.8\sigma_C$ with the MSE function and $\eta_P=5.0$, $\eta_C=0.4$ (note that these are the phase points where we saw the models perform most inconsistently). Additionally, in panels c) and d) we repeat this procedure, and pick the best and the worst radial distribution for respectively the same phase points as panels a) and b), but now obtained with the larger NN models. Comparison of the figures clearly indicates the performance of the larger network leads to a significantly more consistent performance.  To further demonstrate this, we show results for a wide variety of polymer packing fractions in the Appendix (Figs.\ \ref{fig:resultspaircor04smallNNloss}, \ref{fig:resultspaircor04bigNNloss}, \ref{fig:resultspaircor08smallNNloss} and \ref{fig:resultspaircor08bigNNloss}). 

\textit{Pearson Correlation:} Another important observation from this work related our use of the Pearson correlation to evaluate the model.  Interestingly, we found that a high Pearson correlation between the predicted the true free area alone was not sufficient to guarantee accurately performing models. To demonstrate this, in Fig. \ref{fig:comparisonnetwork} we indicate on each plot the Pearson correlations $\rho$ between the predicted and the true free area for the depicted models. As we can see, even the bad performing models are associated with Pearson correlations above $\rho>0.99999$. This underscores the importance of testing the model's predictive performance not only against average quantities, like the Pearson correlation, but also against single particle properties. This testing can be performed either through post-processing analysis, as we do here, or on-the-fly testing conducted during the simulation \cite{li2015molecular, jacobsen2018fly, jinnouchi2019phase}.

\section{Conclusion and outlook}
In conclusion, we developed a physically inspired method that is able to consistently predict the effective many-body potential between colloids in a system of hard colloids and ideal polymers. By describing the local environment of each colloid in terms of its associated Voronoi cell and using a simple neural network to predict the potential energy of this cell, we successfully reproduced the structural properties of the system as well as the effective potential energy and the virial pressure. Here, we validated the accuracy of our methodology by comparison to brute-force simulations.

The main goal of this paper was to develop a new ML strategy for fitting many-body potentials, rather than fully optimizing the strategy. However, depending on the polymer size and packing fraction, we already observed a significant speed up in simulation time between the ML-system and the brute-force system. We believe that further optimizations of the algorithm could enhance this speedup even more.

Due to the simplicity of the benchmark model used in this paper, we were able to explore the training process of the ML potential in more detail. In particular, we found that for this system the ML potential was sensitive to the training data. It turned out that using the configurations of the brute-force system leads to an imbalance in the training data, which stemmed from the fact that the crystal environments were oversampled. 

We also saw that high Pearson correlations between predicted and true free area alone were insufficient to guarantee high accuracy performance. This result underscores the general importance of evaluating a ML model's predictive performance against true values -- not only globally but also on the single particle level. This can be done either via post-processing analysis or on-the-fly testing conducted during the simulation \cite{li2015molecular, jacobsen2018fly, jinnouchi2019phase}.

Our developed methodology is targeted at systems where the effective potential is short-ranged, i.e.\ where most many-body effects are captured in single Voronoi-cell expansions. As a result, we expect that this technique can be extended to other systems with short range potentials, such as colloid-polymer mixtures with non-ideal polymers or systems governed by steric interactions. This approach could also be applied to self-propelled Voronoi models for tissue mechanics\cite{bi2016motility, barton2017active}, where the computational bottleneck is the calculation of the local $n-$body forces, making it possible to simulate, e.g.\ the entire early-stage embryo with several tens of thousands of cells. Additionally, the methodology could also be adapted to 3D. Although most techniques presented in this paper are readily extendable to 3D, implementation of the dynamically updating Voronoi algorithm might be challenging: To our knowledge, there currently exists no algorithm equivalent to equiangulation algorithm in 3D.

\section*{Acknowledgements}
The authors would like to thank Willem Gispen and Marjolein de Jager for many useful
discussions. 

\section*{Data Availability Statement}
All simulation codes needed to reproduce the data, as well as all relevant simulation data and notebooks to analyze the data and generate the figures is published at Zenodo prior to publication of the paper: \url{http://doi.org/10.5281/zenodo.15356418} .

\section{References}
\addcontentsline{toc}{part}{Bibliography}
\markboth{\MakeUppercase{Bibliography}}{}
\bibliographystyle{abbrvunsrt2}
\bibliography{refs}

\appendix
\section{Delaunay- and Voronoi tessellation}
The dynamic implementation of the Voronoi/Delaunay tessellation that we use in this paper is based on the algorithm discussed in Ref.\ \onlinecite{barton2017active}. In this section, we briefly describe how the algorithm works and then discuss the alterations we applied to make it suitable for our system.\\ 
\begin{figure*}[t]
    \centering\includegraphics[width=0.8\textwidth]{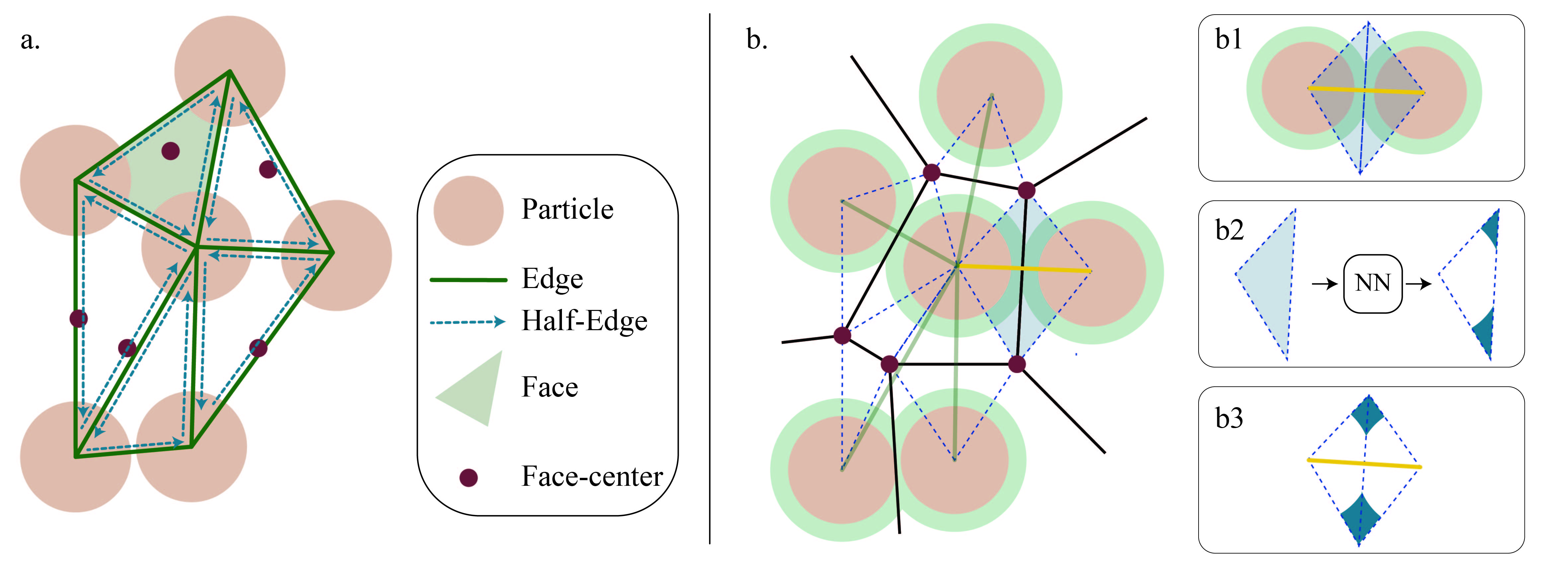}
    \caption{a) Cartoon of the representation of a Delaunay tessellation in the Monte Carlo simulation. Note that the red `face-centers' indicate the centers of the circumscribed circles associated with each triangle. b) Cartoon of a free area prediction. First the Delaunay tessellation (green opaque lines) is converted to a Voronoi tessellation (black lines). Subsequently, we take one of the rhombi consisting of two adjacent triangles that make up the Voronoi cell (b1). We then use the NN to make a prediction for the free area associated with one of the triangles (b2). Finally, the total free area of the rhombus is given by twice the predicted free area (b3). }
    \label{fig:Voronoitri}
\end{figure*}
\subsection{Implementing Delaunay tessellation in Monte Carlo simulation}
To implement the dynamical update algorithm of the Delaunay Triangulation (DT) in our Monte Carlo simulation, we use the scheme proposed by Ref. \onlinecite{sknepnek2023generating}. In the simulation we make use of four different types of structures (see Fig. \ref{fig:Voronoitri}): particles, edges, half-edges, and faces. In the following, we describe each of these:
\begin{itemize}
    \item \textbf{Particles}. These particles make up the vertices of the DT (which correspond to the ``center'' of the Voronoi cell). Each particle is linked to one arbitrarily chosen outgoing half-edge connected to that particle.
    \item \textbf{Faces}. The faces  correspond to the triangles that make up the DT. The face centers (i.e. the centers of the circumscribed circle associated with the triangle) correspond to the vertices of the dual Voronoi tessellation. Each face is linked to one arbitrarily chosen half-edge associated with one of the face edges. 
    \item \textbf{Edges}. Edges are undirected connections between two particles. Each edge is associated with a pair of half-edges (pointing in opposite directions) that connect the same two particles. Throughout the simulation, the pair of half-edges and the edge will always be linked to each other.  
    \item \textbf{Half-edges}. In contrast to the edges, the half-edges are directed and point from particle $i$ to particle $j$. Moreover, half-edges are also linked to other half-edges, and  point from the previous and to the next half edge that together enclose a face in the counterclockwise direction.  The reason that we use these directed half-edges is that they allow us to traverse over the tessellation. One can e.g. loop over all particles associated with one triangle, or loop over all neighbouring particles $j$ associated with particle $i$. 
   
\end{itemize}

In the MC simulation, each time a particle is moved we have to update the DT. To converge to the correct DT (and thus VT), we implement the equiangulation procedure\cite{brakke1992surface} as discussed in the main text in the following manner:
\begin{enumerate}
    \item We set up a linked \textit{checklist}, which contains all the edges that potentially have to be flipped. Initially, this list contains all the edges of the triangles connected to the displaced particle (e.g. in Fig.\ \ref{fig:Voronoitri}a, if we would move the middle particle, all depicted edges should be checked). Additionally, we set-up a linked \textit{changelist} which contains all the edges that are associated with a changed triangle. Note that at the beginning the checklist and changelist are the same.
    \item We iterate over the checklist and for each edge determine if it needs to be flipped (see Fig. \ref{fig:flipmove} and section \ref{sec:MCimplementation}). If so, we flip it,  and add the other four edges that make up the connected triangles to the checklist and the changelist (if they were not already in these lists). We then remove the checked edge from the checklist (whether flipped or not). This process continues until there are no remaining edges in the checklist. At this point, we have reached the correct DT. 
    \item We construct the corresponding VT by computing the new face centers for the triangles that are associated with each edge in the changelist.
\end{enumerate}

In our MC simulation, particle moves are accepted and rejected according to the Boltzmann weight of the difference in free area before and after the move. Note that this assumes that the move does not lead to an overlap between colloids. In the case of an overlap, the move is rejected immediately and no new tessellation is computed. As discussed in the main text, the amount of free area is predicted using a neural network. To calculate the difference in free area between two configurations, we consider the free-area difference between Voronoi triangles that are associated with edges in the changelist (since these are the only Voronoi triangles for which the free area has changed). Moreover, due to the construction of the VT, two adjacent Voronoi triangles are equal, meaning that we only have to apply the NN to one of them, see Fig.\ \ref{fig:Voronoitri} (b3). 
In order to make the algorithm as efficient as possible, we keep a copy of the DT from before the particle move. If the move is not accepted we reset the DT to the tessellation before the move by reverting all edges that changed. 

\subsection{Specifications and alterations}
\subsubsection{Periodic boundaries}
In our system, we use periodic boundary conditions. This has the advantage that the number of triangles in the DT and thus the number of edges, half-edges and faces stays constant. This does, however, imply that one has to be very careful with which periodic image one uses. Sometimes, in systems with periodic boundaries, one uses nearest-image convention. However, whenever the triangle sides become larger than half the box length, the nearest- image convention no longer holds. To solve this problem, each half-edge carries two numbers which keeps track of which periodic image one should take for that triangle edge, for both the $x$ and $y$ direction. When this number is zero, the half-edge does not cross the boundary of the box, while if it is  $\pm1$ it crosses the box in the positive or negative direction. Note that for computing overlap between colloids we use nearest-image convention, as our box is always sufficiently large to allow for this. 

\subsubsection{Forbidden tesselations}
Although the dynamical updating works for almost all instances, one encounters a problem when a particle move leads to a tessellation that  no longer tiles space. This happens when a particle moves in such a way that it crosses one or multiple edges of the previous DT. Since our particles only move by a small amount, in general such an event is very rare. However, it does occasionally happen when triangles are very stretched due to e.g. a phase separation between colloids and polymers. When a particle move leads to a forbidden tessellation, one can solve the problem by breaking up the particle move into smaller sections, where after each step the DT is updated. In our simulation, we consecutively break up the move into $2^n$ steps, where $n = 1, 2, 3,\dots$ until no edges of the DT are crossed. 

\section{Different network size and loss function}
\begin{figure*}
    \centering
    \includegraphics[width=0.8\linewidth]{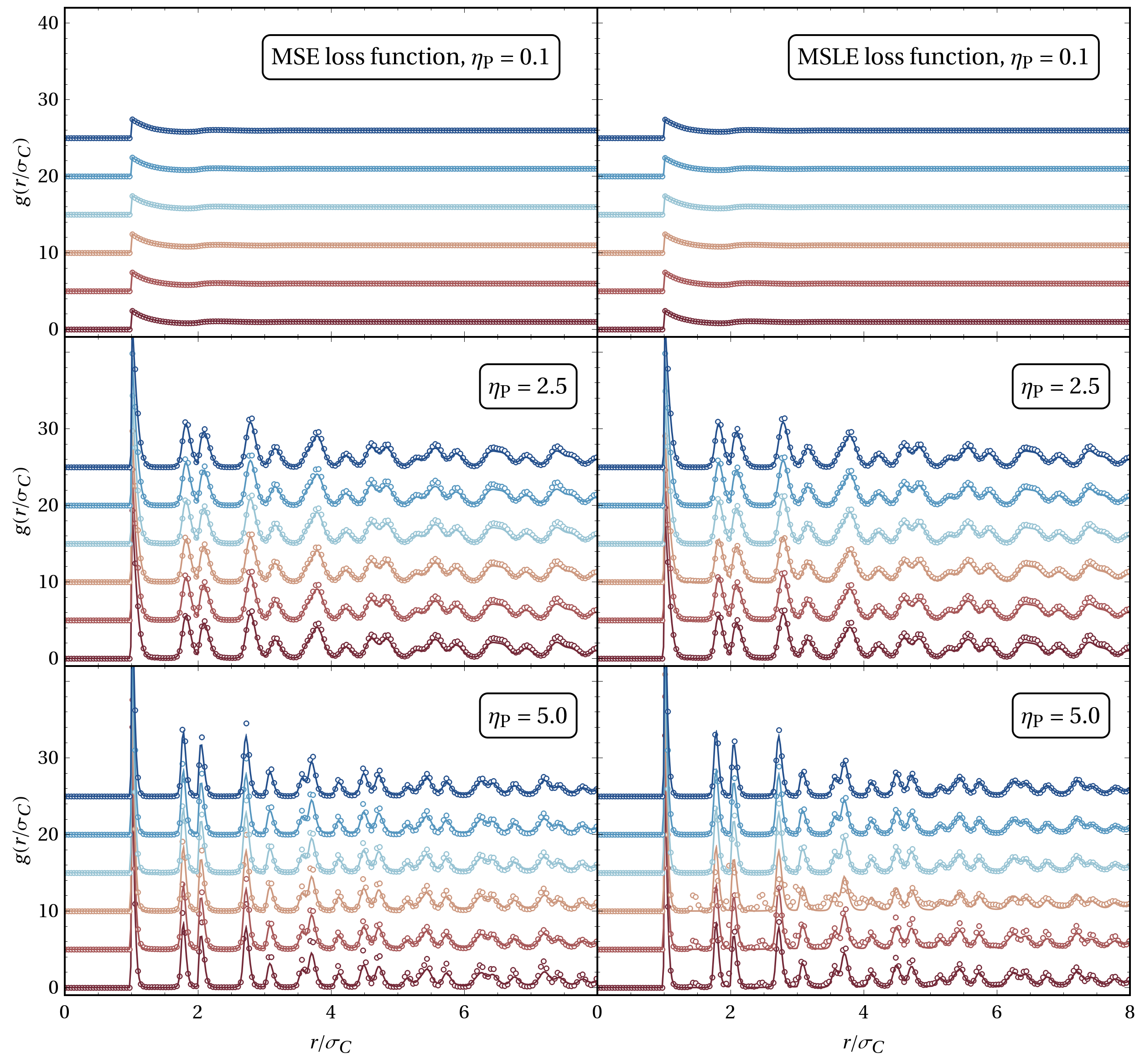}
    \caption{The radial distribution functions as obtained with six independently trained models, for $\sigma_P=0.4\sigma_C$ and respectively an MSE loss function (left panels) and an MSLE loss functions (right panels). Results are obtained with a NN with 3 hidden layers with respectively $[3,3,3]$ nodes. From top to bottom the panels contain results for increasing polymer densities $\eta_P$, where $\eta_P\in[0.1, 2.5,5.0]$. The solid lines represent the ground truth as measured in the system where polymers are treated explicitly, whereas the plotmarkers indicate the radial distribution functions as obtained in the ML system. Both the solid lines as well as the plot markers are averaged over five different simulations. For clarity, different pair correlations are shifted vertically. }
    \label{fig:resultspaircor04smallNNloss}
\end{figure*}

\begin{figure*}
    \centering
 \includegraphics[width=0.8\linewidth]{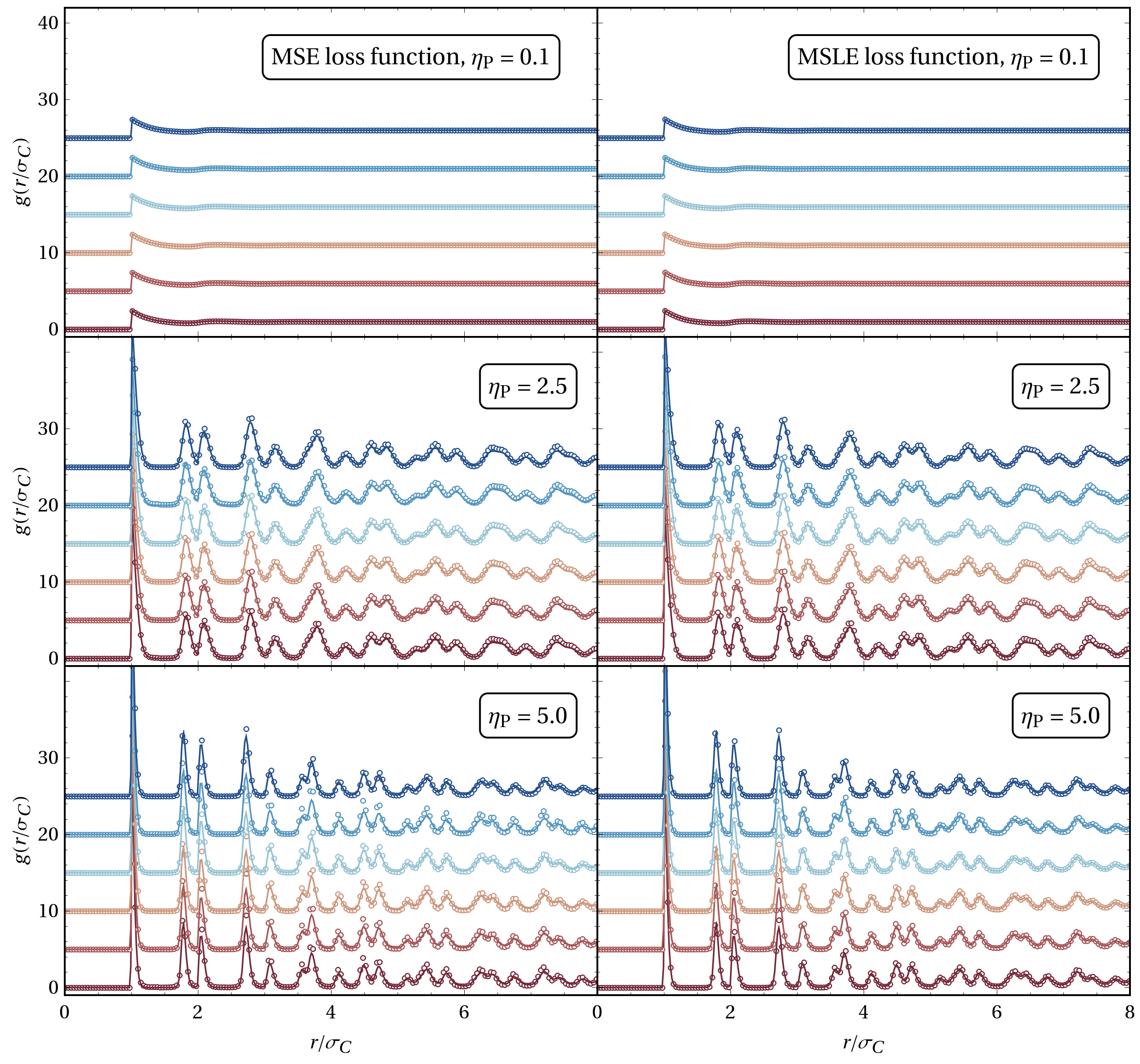}
    \caption{The radial distribution functions as obtained with six independently trained models, for $\sigma_P=0.4\sigma_C$ and respectively an MSE loss function (left panels) and an MSLE loss functions (right panels). Results are obtained with a NN with 4 hidden layers with respectively $[5,5,3,3]$ nodes. From top to bottom the panels contain results for increasing polymer densities $\eta_P$, where $\eta_P\in[0.1, 2.5,5.0]$. The solid lines represent the ground truth as measured in the system where polymers are treated explicitly, whereas the plotmarkers indicate the radial distribution functions as obtained in the ML system. Both the solid lines as well as the plot markers are averaged over five different simulations. For clarity, different pair correlations are shifted vertically. }
    \label{fig:resultspaircor04bigNNloss}
\end{figure*}

\begin{figure*}
    \centering
    \includegraphics[width=0.8\linewidth]{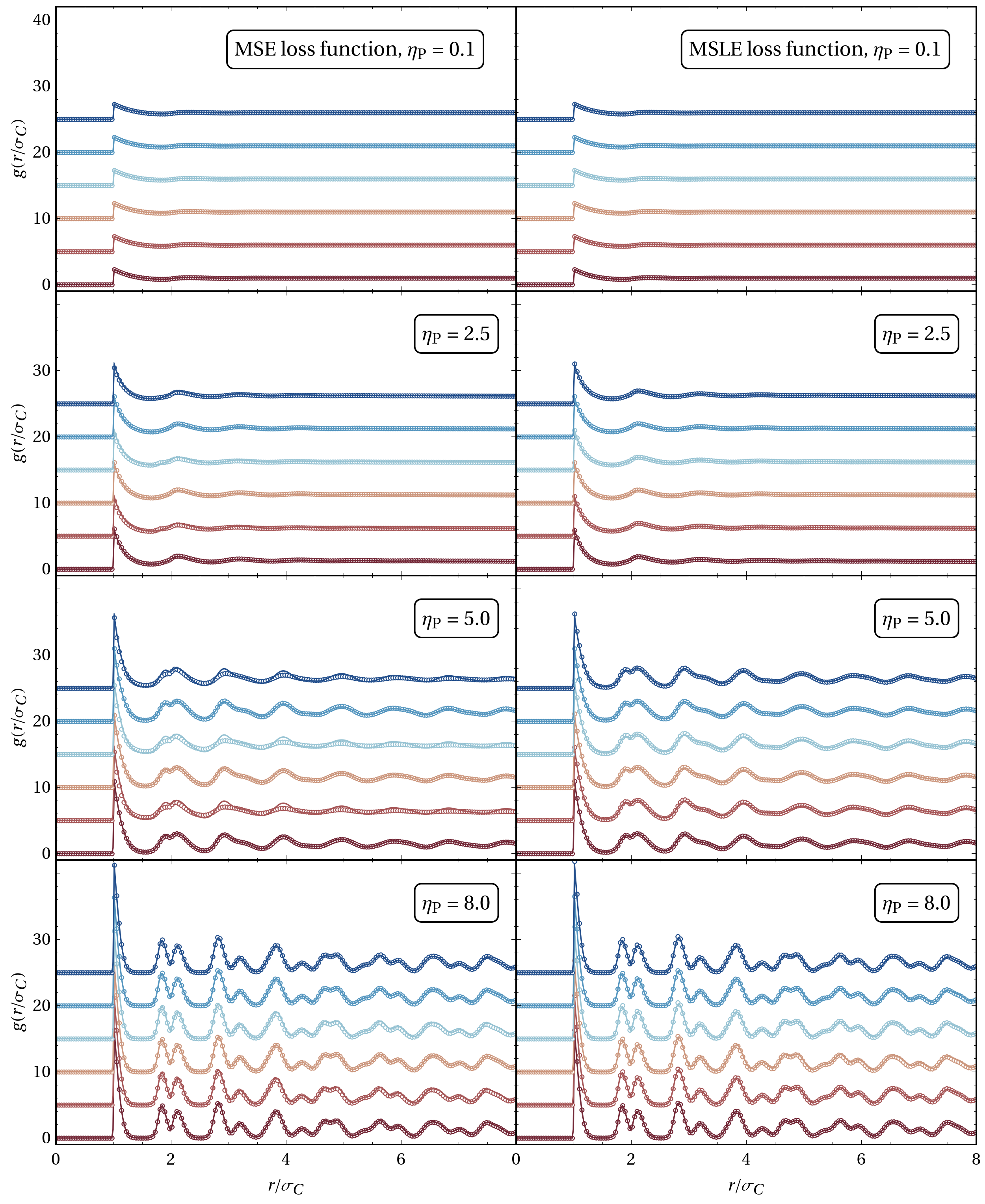}
    \caption{The radial distribution functions as obtained with six independently trained models, for $\sigma_P=0.8\sigma_C$ and respectively an MSE loss function (left panels) and an MSLE loss functions (right panels). Results are obtained with a NN with 3 hidden layers with respectively $[3,3,3]$ nodes. From top to bottom the panels contain results for increasing polymer densities $\eta_P$, where $\eta_P\in[0.1, 2.5,5.0, 8.0]$. The solid lines represent the ground truth as measured in the system where polymers are treated explicitly, whereas the plotmarkers indicate the radial distribution functions as obtained in the ML system. Both the solid lines as well as the plot markers are averaged over five different simulations.  For clarity, different pair correlations are shifted vertically. }
    \label{fig:resultspaircor08smallNNloss}
\end{figure*}

\begin{figure*}
    \centering
 \includegraphics[width=0.8\linewidth]{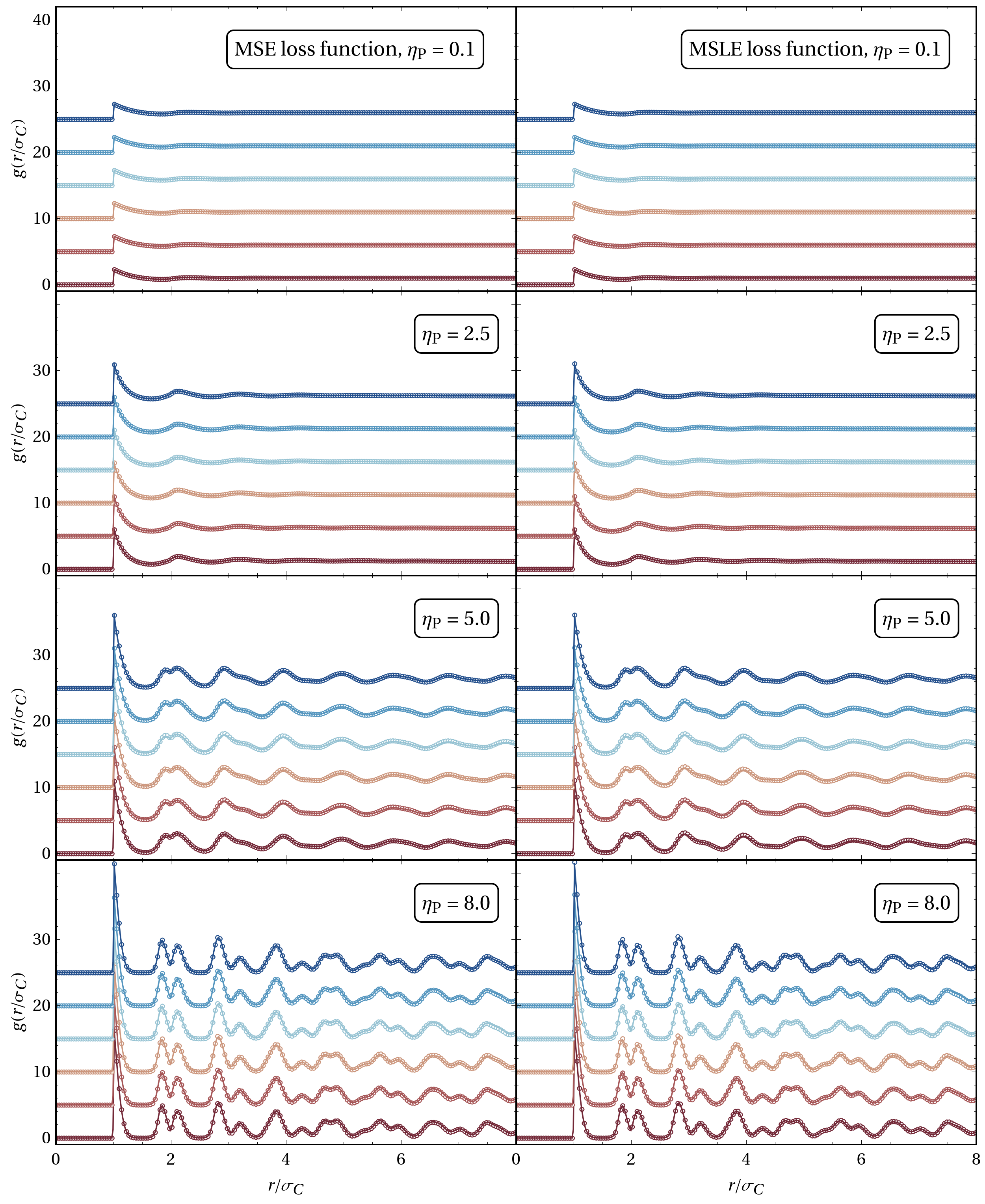}
    \caption{The radial distribution functions as obtained with six independently trained models, for $\sigma_P=0.8\sigma_C$ and respectively an MSE loss function (left panels) and an MSLE loss functions (right panels). Results are obtained with a NN with 4 hidden layers with respectively $[5,5,3,3]$ nodes. From top to bottom the panels contain results for increasing polymer densities $\eta_P$, where $\eta_P\in[0.1, 2.5,5.0, 8.0]$. The solid lines represent the ground truth as measured in the system where polymers are treated explicitly, whereas the plotmarkers indicate the radial distribution functions as obtained in the ML system. Both the solid lines as well as the plot markers are averaged over five different simulations. For clarity, different pair correlations are shifted vertically.}
    \label{fig:resultspaircor08bigNNloss}
\end{figure*}

In this paper, we explored the effect of two different loss functions (MSE and MSLE) and two different network sizes (with 3 and 4 hidden layers respectively) on the accuracy of the ML potential. For each combination of network size and loss function, we trained 6 independent models for two polymer sizes $\sigma_P = 0.4\sigma_C$ and $\sigma_P = 0.8\sigma_C$. In Figs.\ \ref{fig:resultspaircor04smallNNloss} and \ref{fig:resultspaircor04bigNNloss} we plot the colloid-colloid radial distribution functions for $\sigma_P = 0.4\sigma_C$ obtained with the small and big network respectively. Subsequently, in Figs.\ \ref{fig:resultspaircor08smallNNloss} and \ref{fig:resultspaircor08bigNNloss} we plot the colloid-colloid radial distribution functions for $\sigma_P = 0.8\sigma_C$ obtained with the small and big network respectively. For all models, we considered a range of different polymer packing fractions. In the figures, panels on the left display results obtained with models trained with the MSE loss function, while the results in the right panels are obtained with models that were trained with the MSLE loss function. In all figures, solid lines represent the radial distribution functions as measured in the full system, where each line is averaged over five independent simulations. The plot markers correspond to the radial distribution function as measured in the ML simulations, where we again averaged over five independent simulations. 

\end{document}